\documentclass [a4paper, 11pt, titlepage, fleqn,openright, twoside]{book}

\usepackage{fancyhdr}
	\renewcommand{\sectionmark}[1]{\markright{\thesection\ #1}}
	
\usepackage[Glenn]{fncychap}
	\ChRuleWidth{1pt}
	\ChNameVar{\Large\sf}
	\ChTitleVar{\Large\bfseries\rm}

\usepackage [pdftex]{graphicx, color}  
\usepackage [italian,english]{babel}
\usepackage [applemac]{inputenc}
\usepackage{newlfont}
\usepackage{multirow}
\usepackage {indentfirst}
\usepackage{graphicx}
\usepackage{textcomp}
\usepackage{changepage}
\usepackage{amssymb} 
\usepackage{enumerate}
\usepackage{rotating}
\usepackage{caption} 
\usepackage{listings} 
\usepackage{epigraph}
\usepackage[usenames,dvipsnames,svgnames,table]{xcolor} 
\usepackage{hyperref}
\hypersetup{
 citecolor=blue}

\begin{document}

\begin{titlepage}
\changetext{}{85pt}{}{0pt}{}
\begin{center}
{{\Large{\textsc{Rochester Institute of Technology}}}} \rule[0.1cm]{15.8cm}{0.1mm}
\rule[0.5cm]{15.8cm}{0.6mm}
{\small{\bf COLLEGE OF SCIENCE\\
School of Physics and Astronomy\\}}
\end{center}
\vspace{15mm}
\begin{center}
{\LARGE{\bf Reduction of Integral Field Spectroscopic Data from the Gemini Multi-Object Spectrograph}}\\
\vspace{3mm}
{\LARGE{\bf (a commented example)}}\\
\vspace{15mm} {\large{}}
{{\bf $-\cdot$ Davide Lena $\cdot-$}}
\end{center}
\vspace{23mm}
\par
\noindent

\begin{figure}[!hbp]
\centering
\includegraphics[trim = 1cm 1cm 1cm 0cm, clip = true, scale = 0.6]{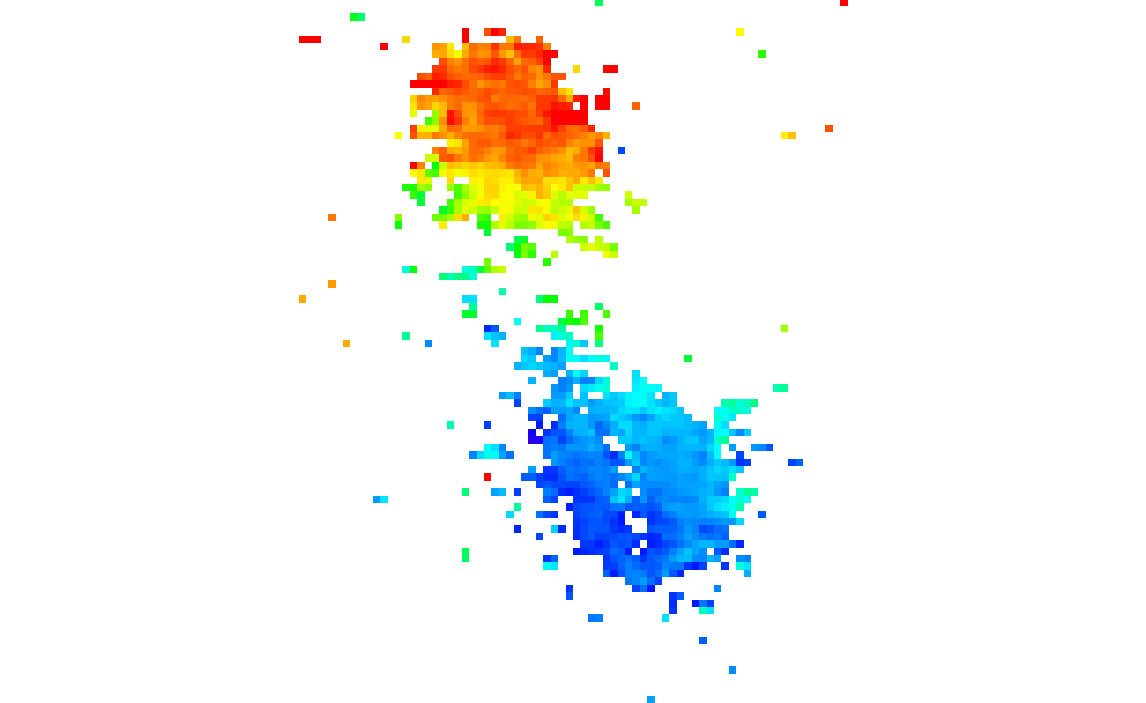}
\end{figure}

\vspace{12mm}
\begin{center}

\rule[0.1cm]{15.8cm}{0.1mm}
{\large{\bf Fall 2014 }}
\end{center}
\clearpage{\pagestyle{empty}\cleardoublepage}
\end{titlepage}

\frontmatter

\chapter{Acknowledgements}
This work started during a visit at the Universidade Federal do Rio Grande do Soul (UFRGS, Brazil), October 2011, while a graduate student at the Rochester Institute of Technology (NY, USA). Special thanks go to Dr. Allan Schnorr-M\"{u}ller for his valuable help in safely guiding me through the tricks and traps of the data reduction procedures. I am grateful to Dr. Rogemar Riffel, who is always willing to share with me his deep knowledge of integral field spectroscopy. I thank Trent Seelig and Sravani Vaddi for testing a previous version of this guide, giving suggestions for improvement and providing some of the figures. Thanks to Dmitry Vorobiev and Dr. Michael Richmond for proofreading the current version providing detailed suggestions to improve the text. I am grateful to my supervisor, Dr. Andrew Robinson, and Dr. Thaisa Storchi-Bergmann for organizing my visit at UFRGS. Last, but not least, I would like to thank the GEMINI staff for their support through the many, many stages involved in the process of turning a ghostly swarm of photons into useful data. 

Financial support was through the NSF grant n AST - 1108786.  
\clearpage{\pagestyle{empty}\cleardoublepage}

\tableofcontents 	
\listoffigures		
\addcontentsline{toc}{chapter}{List of figures}

\clearpage{\pagestyle{empty}\cleardoublepage}

	\mainmatter
	\pagestyle{fancy}  
	
	\fancyhead[RE]{\bfseries\footnotesize\nouppercase{Introduction}}
	\fancyhead[LO]{\bfseries\footnotesize\nouppercase{Introduction}}
	\chapter *{Introduction}
\addcontentsline{toc}{chapter}{Introduction}

This is a commented IRAF script used to reduce data from a galaxy observed with the Gemini Multi-Object Spectrograph Integral Field Unit (GMOS IFU) on GEMINI-South in 2 slits mode. The command list has been adapted from scripts given as tutorials at the South American Gemini Data Workshop (S\~{a}o Jos dos Campos, Brazil, October 27-30, 2011) and scripts kindly provided by Dr. Allan Schnorr-M\"{u}ller, at that time graduate student at the Universidade Federal do Rio Grande do Soul. 

In \textsection \ref{chap: suggestions} general advice and comments on the data reduction procedure are given; in \textsection \ref{chap: star} I present a commented reduction of the standard star; in \textsection \ref{chap: science} the reduction process is applied to a galaxy (note that some steps are equivalent to the standard star reduction process, in those cases see \textsection \ref{chap: star} for full comments). Typical problems encountered in the reduction process are discussed in appendix \ref{app1}; some useful definitions are given in appendix \ref{app: def} and suggested readings are listed in appendix \ref{sec: reading}.

Feedback is welcome and encouraged. Contact the author for the most recent version of this document.
\vskip10pt

e-mail: \color{BrickRed}{\href{mailto:dxl1840@g.rit.edu}{\textbf{dxl1840 at g.rit.edu}}}\color{Black}{}
	\clearpage{\pagestyle{empty}\cleardoublepage}
	
	\fancyhead[RE]{\bfseries\footnotesize\nouppercase{\leftmark}}
	\fancyhead[LO]{\bfseries\footnotesize\nouppercase{\rightmark}}

\chapter{Before starting}
\label{chap: suggestions}

\section{Organize your files}
\begin{itemize}
\item processed
\item raw
\end{itemize}
The folder \textit{raw} contains all of the raw files. 
If observations have been taken over multiple nights, then it is a good idea to make separate reductions for each night (as every night has its own flats and arcs) and, for easy reference, it is a good idea to create the folder:

\begin{itemize}
\item raw$\_$backup
	\begin{itemize}
		\item night 1
			\begin{itemize}
			\item calibration: daytime calibration raw images (bias, arc, flat)
			\item standards: nighttime calibration raw images (standard stars)
			\item science: raw images of your target
			\end{itemize}
		\item night 2 
			\begin{itemize}
			\item $\ldots$
			\end{itemize}
	\end{itemize}
\end{itemize}
For practical purposes related to the reduction pipeline, it is easier to work in the folder \textit{raw}, which contains all of the raw data. 

The observation log will be useful to identify the nature of the file content and send it to the appropriate folder (i.e. calibration, standards, science). 
In the log file, look for a column labelled \textbf{File Number}, this gives the last 3 digits of the file name. Under \textbf{Target Name}, the label assigned to the target can be found, e.g. M31 (the main object to be observed), GCALflat (the flat), CuAr (the arc or lamp), star name (the calibration star, e.g. LTT3218). Essential are the labels in the column \textbf{Disperser}, under which important information about the nature of the observation is given:
\begin{description}
\item[mirror:] indicates that the file is an \textit{acquisition}, just a quick snapshot of the main target used to accurately place it in the field of view of the camera, the slit or the integral field unit of a spectrograph.
\item [e.g. R400/650:] indicates the disperser (in this case R400) and central wavelength of the filter used to make the observations (in this case 650 nm); often more than one central wavelength is used (e.g. if spectral dithering was applied there will be a set of observations performed at 650 nm and a set at 655 nm). 
\end{description}

Information about the file content is also stored in the file header which can be accessed, e.g., from the IRAF\footnote{IRAF is the Image Reduction and Analysis Facility, a general purpose software system for the reduction and analysis of astronomical data. IRAF is written and supported by the National Optical Astronomy Observatories (NOAO) in Tucson, Arizona.} terminal with the commands: \textit{imhead *[0]} or, after loading the packages \textit{gemini}, \textit{gemtools}, with \textit{gemextn filename.fits}.

Note: this reduction process is time consuming (hours). In addition, especially when working with GEMINI-South data, there can be tricky problems to fix. If the user is not an expert, the problem may be evident only at the very last step of the reduction process. Hence, it is an excellent idea to start applying the entire reduction process on the standard star. This requires a reasonable amount of time to be completed (about two hours) and allows one to see if there are any major problems which need to be taken care of. 
\vskip10pt

Final suggestion before starting: always, always visually inspect the data (before, during and after the reduction process).

\section{A note on the overscan}
\label{sec: overscan}
Processed bias files given in the GEMINI archive come with no overscan subtraction (this was decided early in the history of GEMINI and for consistency it has not been changed). 
Moreover, the overscan is trimmed therefore, if we want to subtract the overscan from the bias, we have to find individual bias frames\footnote{See: \url{http://www.gemini.edu/sciops/data-and-results/science-archive/calibration-data-retrieval/instructions}} and process them, subtracting the overscan.

If the bias to be used in the data reduction is not overscan subtracted, then the overscan should not be subtracted from the data. Subtracting a bias not overscan-subtracted from overscan-subtracted data will subtract too much from the data and, e.g., the flux of the source will be underestimated.
 
Throughout this data reduction example we will use a bias that was not overscan-subtracted, therefore we will not overscan-subtract the data.

\clearpage{\pagestyle{empty}\cleardoublepage}

\chapter{The Standard Star}			         
\label{chap: star}
Start with the reduction of the standard star. Why? First, as the whole reduction pipeline is quite time consuming (hours in the best case), reducing just the star may help to identify in a shorter amount of time any instrumental problem which needs to be fixed (e.g. fiber misidentification, use of the wrong flat or bias). Once the problem is fixed on the standard star, it is likely (but not assured) that the reduction of the science data will not present any further problems. Second, the standard star will be used to calibrate the flux of the galaxy.

\section{Prepare IRAF} 
\label{sec: prepareIRAF}  
Assuming that all of the raw files are in the folder \textit{raw} and the processed files will be sent to the folder \textit{proc} (both folders must have been previously created by the user):
\begin{lstlisting}
1. Define variables:

  set rawdir = ``/Users/...YourPath.../raw/''
  set procdir = ``/Users/...YourPath.../proc/''

2. Load the IRAF packages and set values to default:
  
  gemini
  gemtools
  gmos

  unlearn gemini gemtools gmos

3. Set the logfile:

  gmos.logfile=``Star.log''
\end{lstlisting}

\begin{lstlisting}
4. Set common parameter values for the IRAF task gfreduce:

  gfreduce.rawpath=``rawdir$''
  gfreduce.fl_fluxcal=no
\end{lstlisting}

Specify the bias to use:
\begin{lstlisting}
  gfreduce.bias=``rawdir$gS20110923S0301_bias''  
\end{lstlisting}

Note: if an observing program has observations carried out over a long period of time (e.g. two months) then more than one bias file will be associated with the program (probably two for this particular example). Make sure to use the one that was taken close to the observations that are undergoing reduction.

\section{Flat reduction and fibers identification}
In this step, \textit{gfreduce} will be applied to the flats. Run it in interactive mode for the first flat. We have to check if the fibers have been identified correctly. Afterwards, it can be run in non-interactive mode for all the remaining flats.
 \begin{lstlisting}
  gfreduce S20110923S0269 \
  fl_gscrrej- \
  fl_wavtran- \
  fl_skysub- \
  fl_inter+ \
  fl_over- \
  slits=both
  \end{lstlisting}
  
  where \textit{S20110923S0269.fits} is the flat. Note that we set \textit{fl\_over-}, which means that the overscan will not be subtracted from the flat (see \textsection\ref{sec: overscan} for additional comments on overscan subtraction). If the bias in use was overscan-subtracted, then \textit{fl\_over+} should be used here.

Once the above commands are given, the user will go through the following steps.

  \subsection{IRAF questions and actions}
\begin{figure}[p]
\centering
\includegraphics[draft=false, width= \textwidth]{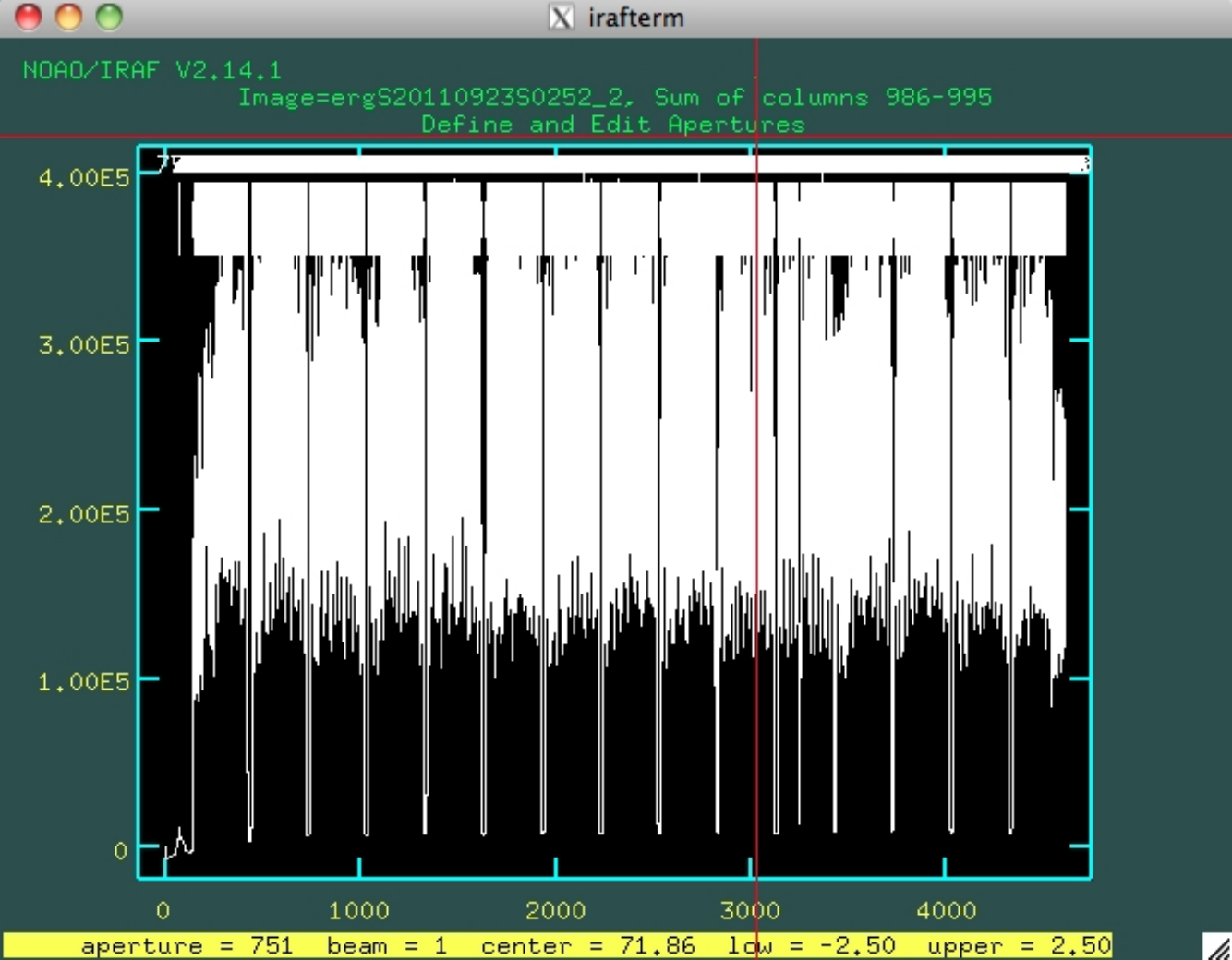}
\caption[IRAF interactive window with fibers blocks]{Interactive window displayed by IRAF when running \textit{gfreduce} in interactive mode. Zoom-in on each fibers block (selecting region by placing the red cross on the bottom left corner of the region to expand and typing ``we" and ``e" on the top right). On top there are numbers identifying each fiber. The first fiber of each block should begin with x1 (1, 51, 101, \ldots) while the last fiber of each block should be identified with x0 (50, 100, 150, \ldots). This may not be the case if the last fiber is missing (therefore the last good fiber of the block would be, e.g., 49) or if there was a fiber misidentificantion (e.g. one fiber was assigned two IDs). Misidentification has to be corrected as explained in appendix \textsection \ref{subsubsec: problems}.} 
\label{fig: FibersID}
\end{figure}

\begin{figure}[p]
\centering
\includegraphics[draft=false, width= \textwidth]{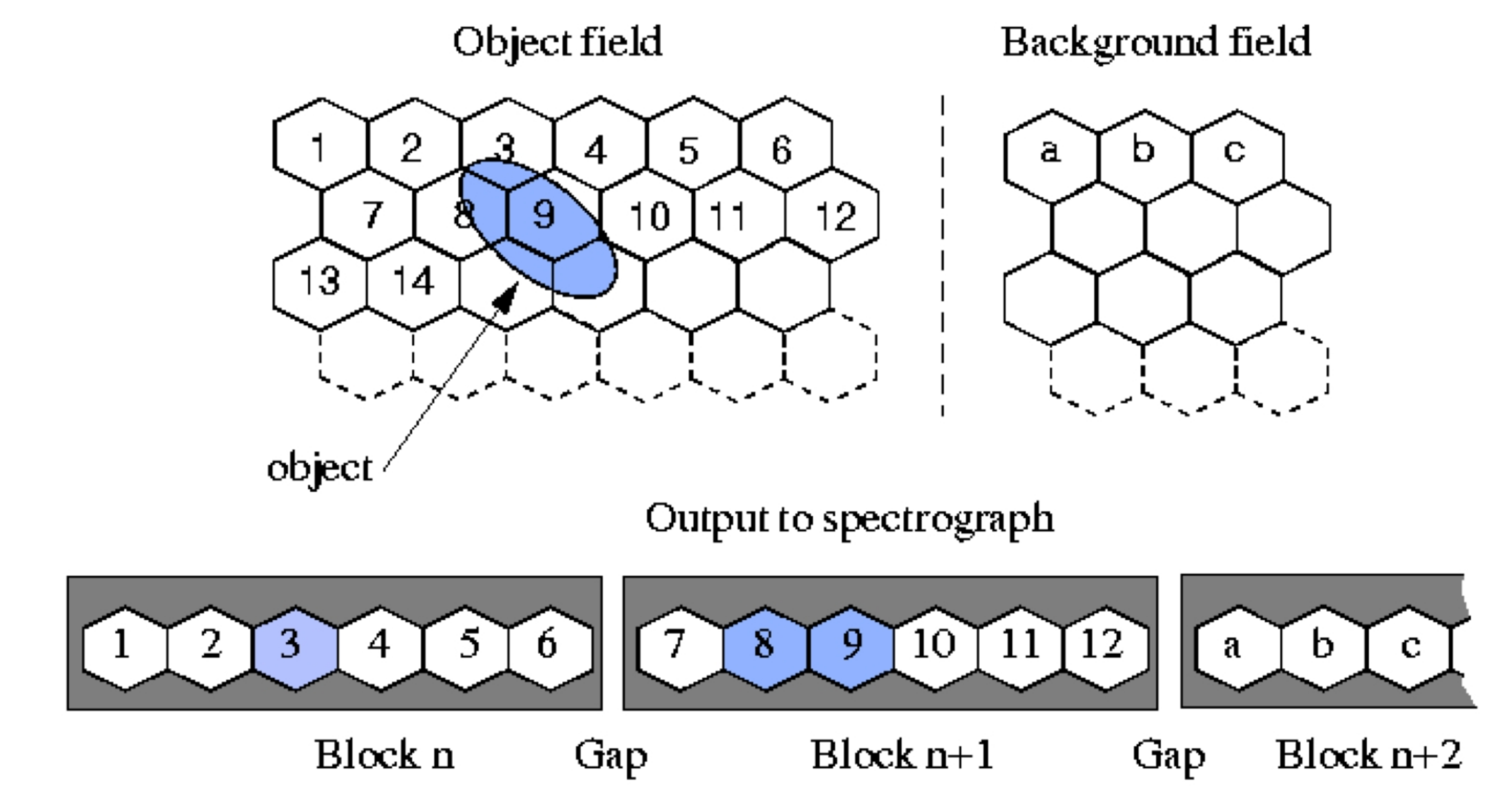}
\caption[Blocks of fibers]{Blocks of fibers. Note that we have blocks of 50 fibers. Credits: image from the ppt presented at the GEMINI data workshop in S\~{a}o Jos Dos Campos (2011) by Rodrigo Carrasco.} 
\label{fig: fibers_groups}
\end{figure}

	  \begin{itemize}
	 \item ``find apertures for ergFileName?" yes
	\item ``recenter apertures for ergFileName?" yes
	\item ``edit apertures for ergFileName?" yes. Here the fibers ID should be checked (this is a very important step): IRAF will open an interactive window as shown in Fig.\ref{fig: FibersID}. Fibers are in groups of 50 because the instrument configuration is similar to what is shown in Fig.\ref{fig: fibers_groups}. Check the ID assigned to the first and the last fiber of each group by zooming-in (select the region by typing ``we" on the bottom left corner of the region to expand and ``e" on the top right, ``wa" will zoom-out). On top of each fiber there is a number, the ID. The first fiber of each group should begin with x1 (1, 51, 101, 151, \ldots) while the last fiber of each block should be identified with x0 (50, 100, 150, \ldots). If this is not the case, it may be due to two reasons: 1. one or more fibers at the end are missing, hence they are not numbered (in this case the last good fiber of the block might be identified with e.g. 49); 2. fibers have been misidentified (e.g. two IDs have been assigned to a single fiber), in this case the identification has to be corrected as explained in appendix \ref{subsubsec: problems}. This is important and must be taken care of, otherwise, when making the datacube near the very end of the process, we might get the error: \\ \textit{gfcube: num image rows != num good fibers in MDF}, or we might obtain a cube showing clear signs of bad reconstruction.
	\item ``trace apertures for ergFileName?" yes
	\item ``fit traced positions interactively?" NO
	\item ``write apertures for ergFileName to database?" yes
	\item ``extract aperture spectra for ergFilename?" yes
	\item ``review extracted spectra?" NO
	\item ``warning, coordinate system ignored. Using pixel coordinates." Don't mind.
	\end{itemize}
  
During this first step we might receive the following warning: ``WARNING - GGAIN: gain and read noise not found in /Users/YourPath/iraf/extern/gemini/gmos/data/gmosamps.dat for this mdd. Using values from parameter". It is likely that the IRAF version in use is getting old and the database in gmosamps.dat is outdated. Updating IRAF (argh!) or just substituting this file from a newer IRAF version should solve the issue. 

Output of this first step is: \textit{ergS20110923S0269.fits}.

\section{Twilight flat reduction}			      
 \begin{lstlisting}
 gfreduce S20110924S0020 \
 fl_wavtran- \
 fl_skysub- \
 fl_inter- \
 trace- \
 ref = ergS20110923S0269 \
 fl_gscrrej- \
 fl_over- \
 slits = both
 \end{lstlisting}
 
 where \textit{S20110924S0020.fits} is the twilight flat and \textit{ergS20110923S0269.fits} is the ``reference" file, i.e. the processed flat where fibers were - hopefully - correctly identified. If the bias in use was overscan-subtracted, then \textit{fl\_over+} should be used here.
 
 Output: \textit{ergS20110924S0020.fits}.

\section{Creation of the response curve}			      		
Using the reduced flat we make the response curve with twilight correction:

\begin{lstlisting}
gfresponse ergS20110923S0269 ergS20110923S0269_resp \
sky= 'ergS20110924S0020' \
order=95 \
fl_inter+ \
func=spline3 \
sample=``*"
\end{lstlisting}

where \textit{ergS20110923S0269.fits} is the processed flat and \textit{ergS20110924S0020} is the reduced twilight flat.

Output: \textit{ergS20110923S0269\_resp.fits}.

\section{Reduction of the arc}
\label{sec: arc_star}
\begin{lstlisting}
gfreduce S20110923S0299 \
fl_wavtran- \
fl_inter- \
ref=ergS20110923S0269 \
recenter- \
trace- \
fl_skysub- \
fl_gscrrej- \
fl_bias- \
fl_over- \
order=1 \
weights=``none"
slits=``both"
\end{lstlisting}

where \textit{S20110923S0299.fits} is the arc. More than one arc might be present if observations were carried out at different central wavelengths. It is essential to chose the arc of the same wavelength of the science data.

Set \textit{fl\_over+} if the bias was overscan-subtracted.

Output: \textit{ergS20110923S0299.fits}.

\section{Establish wavelength calibration}
\label{sec: establish_wc}
An important parameter in determining the outcome of the wavelength calibration is the order of the function used to determine the wavelength calibration. This is specified by the parameter ``order" in \textit{gswavelength}. The default value given by IRAF should be ``order = 4" but it safer to start with ``order = 3" (a higher order function could fit undesirable features causing a bad wavelength calibration).

 \begin{lstlisting}
 gswavelength ergS20110923S0299 \
 fl_inter+ \
 nlost=10 \
 order=3
 \end{lstlisting}
 
First, identify some of the lines shown in the IRAF window: zoom-in on the lines, then put the red cross on the peak and mark it typing ``m", in the plotting window. On the bottom, IRAF will show the position of the cross and the inferred position in parenthesis. Use ``:label coord" to see the identified lines and compare with the appropriate spectrum at \url{http://www.gemini.edu/sciops/instruments/gmos/calibration?q=node/10469}. If values in parenthesis are fine, just press ENTER, then ``f" to fit. Bad points can be deleted with ``d". Then ``q" will complete and quit.

The RMS (in angstroms) of the wavelength calibration will be written in the log file for each aperture. This should be inspected and it is recommended that a value  $\leq$10\% of the spectral pixel size should be attained. 
\vskip5pt

Note: it is important to perform this step in interactive mode (by setting $fl\_inter+$); as stated in the \textit{gswavelength} help page, the automatic line identification is not reliable. This might not be true for more recent versions of IRAF and the reader is encourage the check the help page for the most updated information.

\section{Apply wavelength calibration to the arc}
\label{sec: apply_wc_arc}
 
 \begin{lstlisting}
 gftransform ergS20110923S0299 wavtran=ergS20110923S0299
  \end{lstlisting}
 
(note that $gftransform$ will give an error if the file extension (.fits) is included). 
The resulting file will be \textit{tergS20110923S0299.fits}. Visually inspect the arc after the wavelength calibration is applied: lines should appear as straight as possible. Lines that look broken are indicative of a bad wavelength calibration and a possible cause is the use of an excessively high order for the fitting function. Obviously, if the problem is present at this stage, it will propagate to all data to which this calibration will be applied. See \textsection\ref{sec: poor_wc} for additional comments and examples of poor wavelength calibration.

\begin{figure}[h]
\centering
\includegraphics[draft=false, width= \textwidth, trim = 10cm 8cm 10cm 8cm, clip]{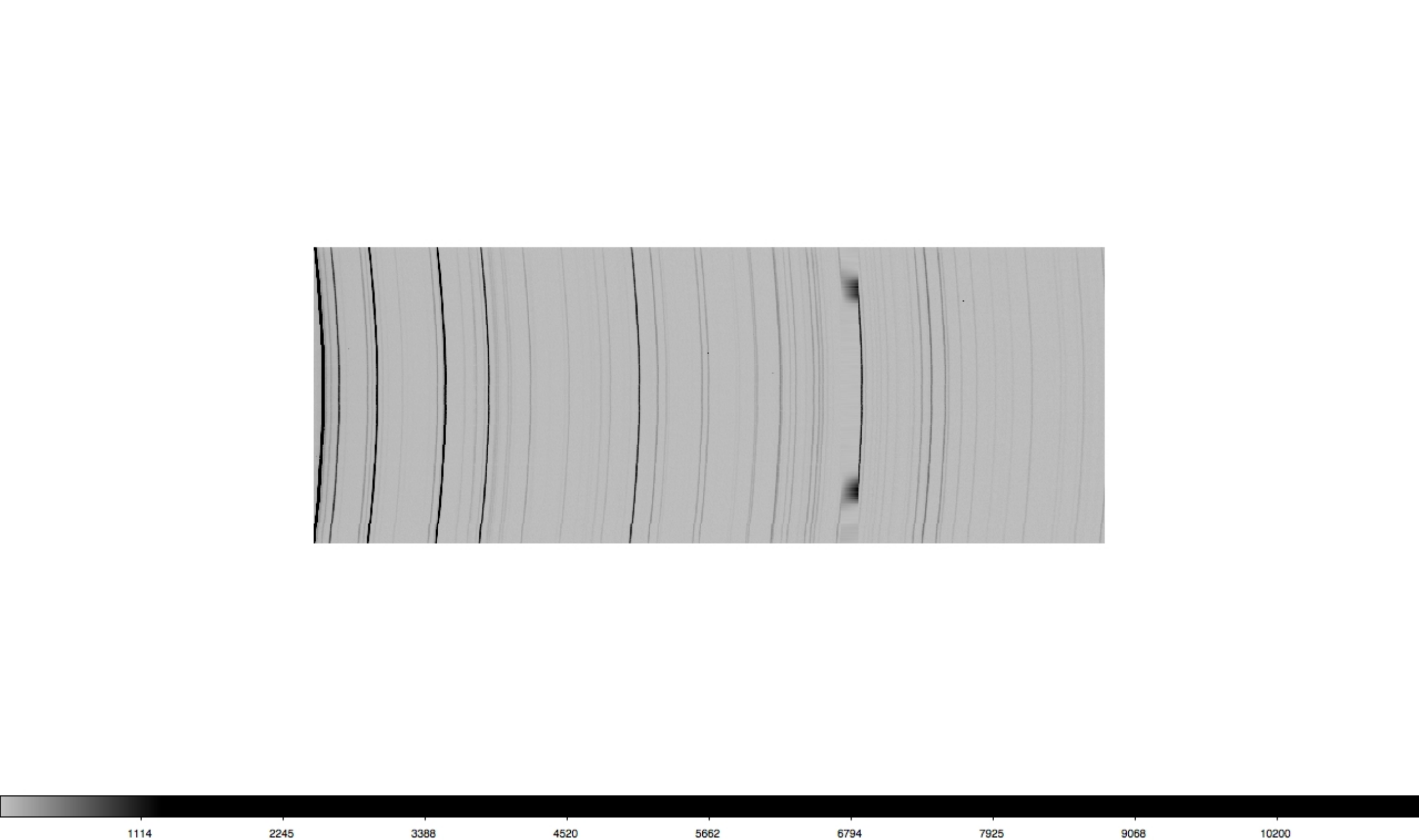}\\
\includegraphics[draft=false, width= \textwidth, trim = 10cm 6cm 10cm 3cm, clip]{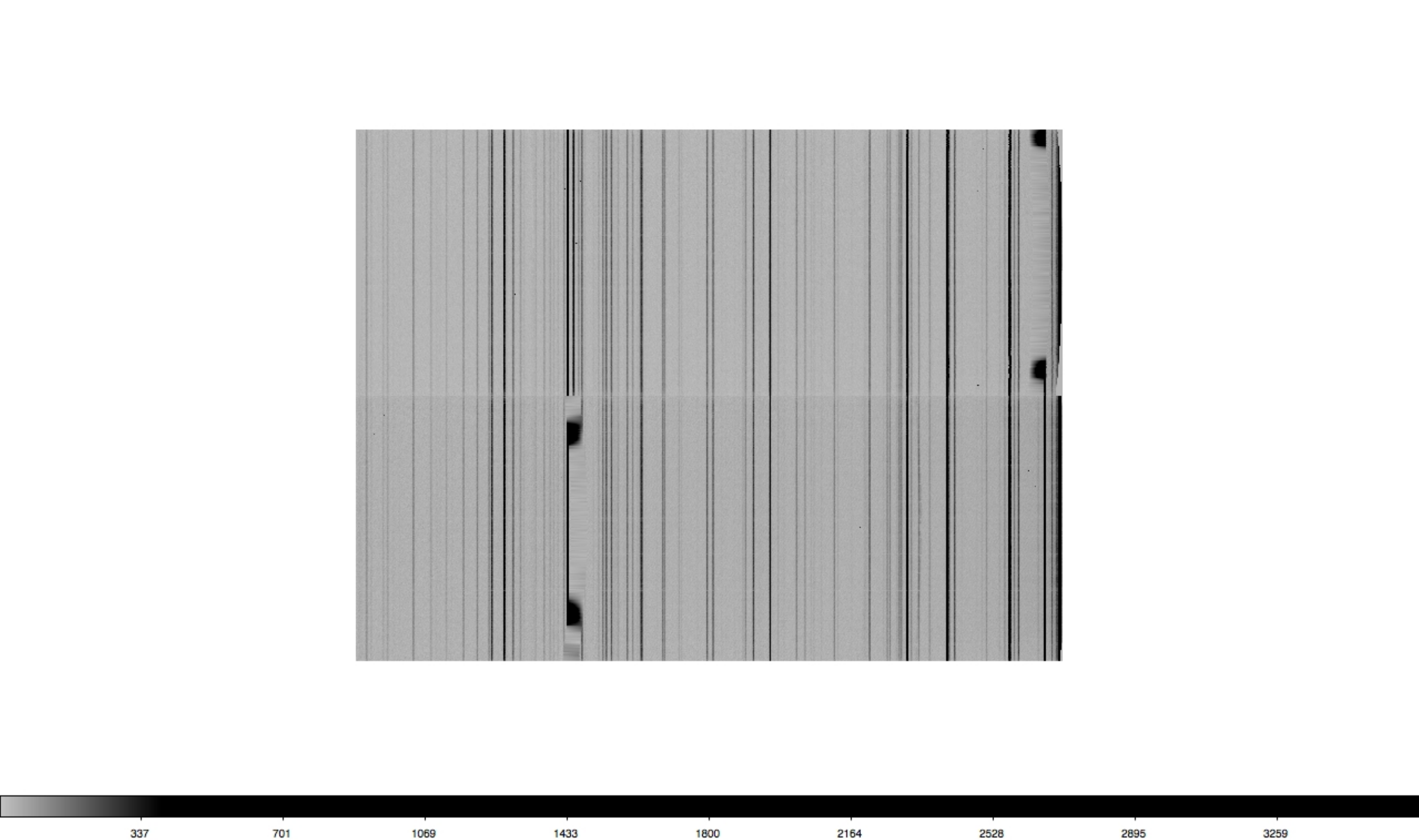}\\
\caption[Arc wavelength calibration]{\textit{Top}: arc before wavelength calibration. \textit{Bottom}: arc after wavelength calibration.} 
\label{fig: arc_calib}
\end{figure}

Fig.\ref{fig: arc_calib} gives an idea of the difference between the arc before and after wavelength calibration.

\section{Reduction of the star} 	 						
 \subsection{Bias subtraction and fiber tracing}
 \label{subsec: gtbs}

\begin{lstlisting}
gfreduce S20110923S0268 \
slits=both \
fl_inter- \
fl_addmdf+ \
key_mdf=MDF \
mdfdir=``rawdir$" \ 
mdffile=gsifu_slits_mdf.fits \
fl_over- \
fl_trim+ \
fl_bias+ \
fl_flux- \
fl_gscrrej- \
fl_extract- \
fl_gsappwave- \
fl_wavtran- \
fl_skysub- \
weights=none
 \end{lstlisting}
 
where \textit{S20110923S0268.fits} is the star. Note: with the parameters \textit{mdfdir} and \textit{mdffile} we specify to use a mdf from the directory rawdir (directory defined in \textsection\ref{sec: prepareIRAF}); this needs to be done only if there were problems with the fibers and the mask definition file (mdf, see Appendix \ref{app: def} for a definition) was modified. Furthermore, we set \textit{weights=none}; as pointed out at \url{http://ifs.wikidot.com/gmos} this produces a better result than \textit{weights=variance}.

Output: \textit{rgS20110923S0268.fits}.

\subsection{Flatfielding and spectra extraction} 
\label{sec: flat_spec}
Here we apply the flat field to the data (via the response curve contained in the file \textit{ergS20110923S0269\_resp.fits}) and we clean the data of cosmic rays:

 \begin{lstlisting}
gfreduce rgS20110923S0268 \
fl_inter- \
fl_addmdf- \
fl_over- \
fl_trim- \
fl_bias- \
fl_gscrrej+ \
fl_extract+ \
fl_wavtran- \
fl_sky- \
fl_flux- \
slit=both \
trace- \
verb+ \
refer=ergS20110923S0269 \
response=ergS20110923S0269_resp \
weights=none
\end{lstlisting}

where \textit{rgS20110923S0268.fits} is the output of \textsection \ref{subsec: gtbs}.

Output: \textit{exrgS20110923S0268.fits}.

\subsection{Wavelength calibration} 
We apply the wavelength calibration on the output of the previous step as follows:

 \begin{lstlisting}
gftransform exrgS20110923S0268 wavtran=ergS20110923S0299
 \end{lstlisting}
 
 where \textit{ergS20110923S0299} is the processed arc. Note, more than one arc might be present if observations have been performed at different wavelengths or on different days. Make sure to use the arc obtained on the same day, with the same grating and filter as those used for the standard star.
 
Output: \textit{texrgS20110923S0268.fits}.
 
\subsection{Sky subtraction} 				
 \begin{lstlisting}
gfreduce texrgS20110923S0268 \
fl_inter- \
fl_addmdf- \
fl_over- \
fl_trim- \
fl_bias- \
fl_gscrrej- \
fl_extract- \
fl_wavtran- \
fl_sky+ \
fl_flux- \
slit=both \
trace- \
verb+ \
weights=none
 \end{lstlisting} 
 
 where \textit{texrgS20110923S0268} is the file produced by \textit{gfreduce rgS20110923S0268.fits} (\textsection\ref{sec: flat_spec}).
 
Output: \textit{stexrgS20110923S0268.fits}.
 
\subsection{Creating the cube} 		
\begin{lstlisting}
 gfcube stexrgS20110923S0268 ssample=0.05
 \end{lstlisting}
 
Output: \textit{dstexrgS20110923S0268.fits}.

\subsection{Making the sensitivity curve} 
\label{subs: sensitivity}
 First, sum all spectra into one:
\begin{lstlisting}
gfapsum stexrgS20110923S0268 \
outimages=``astd" \
combine=``sum"\
reject=``pclip"\
fl_inter-
\end{lstlisting}

Output: \textit{astd}.
 
Now create the sensitivity curve using \textit{gsstandard}. It takes in input spectra from the standard star to be used as calibrators. In this case the name of the file is ``astd" (it was created in the previous step). The output will be a one dimensional image, a simple FIT file, not a MEF (multi extension fits). The name of the output is specified as the argument of \textit{sfunctio}.

\begin{lstlisting}
gsstandard astd \
sfile=``std3"\
sfunctio=``sens" \
starnam=``L1788" \
fl_inte+ \
observa=``Gemini-South" \
functio=``chebyshev" \
order=4 \
caldir=onedstds$ctionewcal/ 
\end{lstlisting}

Output: \textit{sens}.

If \textit{$fl\_inte+$} is set, the process will be interactive. This is strongly recommended. First, we will be asked to edit the bandpasses. This is a very important step because it determines the calibration of the flux, so it is wise to visually inspect what is going on. If something is wrong, the final flux of the object will show unphysical features and we will have to repeat the reduction process. See \ref{app: SF} for the description of a problem that may be encountered.

\begin{figure}[h]
\centering
\includegraphics[trim = 0cm 0cm 0.15cm 0cm, clip = true, draft=false, width= 0.9 \textwidth]{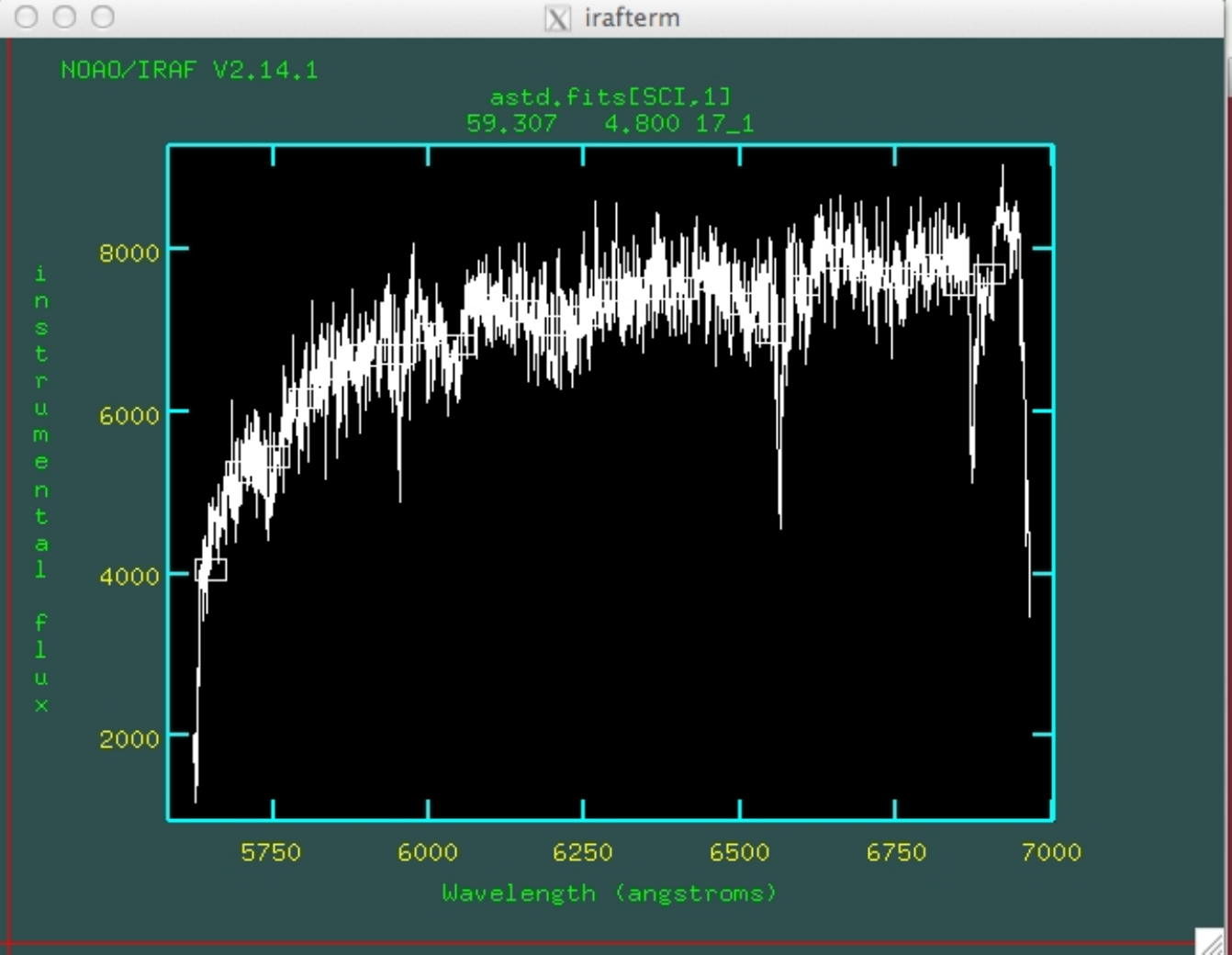}
\caption[Instrumental flux vs wavelength]{Instrumental flux as a function of wavelength. Bandpasses are shown as white boxes on top of the spectrum.} 
\label{fig: bandpasses_correct}
\end{figure}

If we choose to edit the bandpasses, an interactive window similar to Fig.\ref{fig: bandpasses_correct} will open. In this case, there are no clear problems (what is displayed looks like a reasonable stellar spectrum), therefore no actions are necessary. If the spectrum is similar to Fig.\ref{fig: bandpasses_problem}, then something went wrong. In this case the reader should refer to appendix \ref{app: SF} for additional comments. 

\begin{figure}[h]
\centering
\includegraphics[draft=false, width= 0.9 \textwidth]{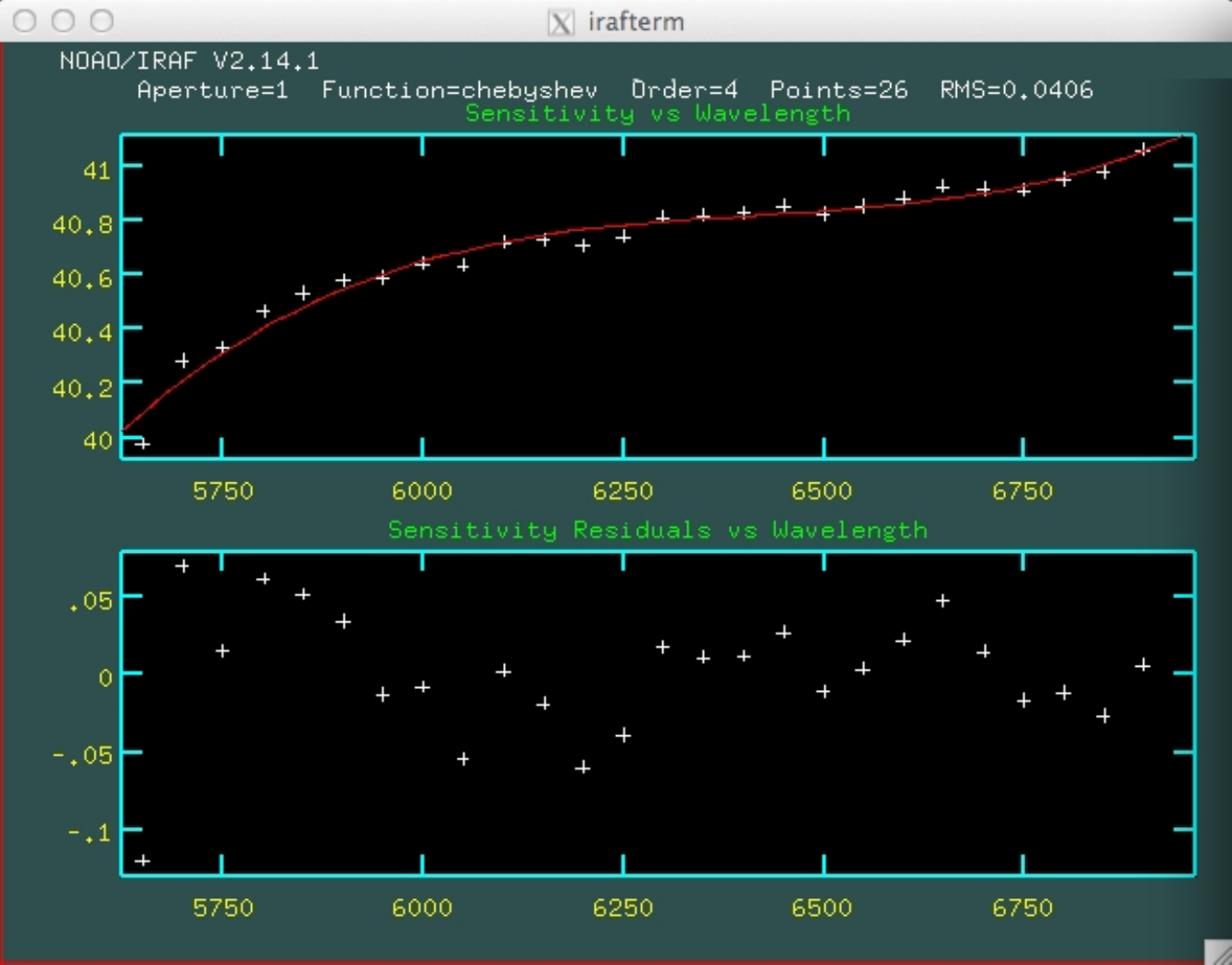}
\caption[Sensitivity curve]{Sensitivity curve.} 
\label{fig: sensitivity_correct}
\end{figure}

Supposing that the spectrum didn't show any problem, now we can quit the interactive display and proceed to the following step. A sensitivity curve similar to the one shown in Fig.\ref{fig: sensitivity_correct} will result. 
At this point IRAF will ask if we want to make the fit interactively. Again, there are no obvious problems, therefore the fit can be run in non-interactive mode.

\subsection{Flux calibration}	 		     
\label{subs: fcal}

Spectra will be divided by the sensitivity curve created in \textsection\ref{subs: sensitivity}:
\begin{lstlisting}
gscalibrate astd \
sfuncti=``sens"\
observa=``Gemini-South" \
fluxsca=1
\end{lstlisting}

Make sure to specify the correct telescope (Gemini-South or Gemini-North). This step completes the reduction of the star.

In order to visualize the cube with, e.g., QFitsView\footnote{QFitsView is a FITS file viewer similar to SAOimage, DS9, etc. In addition, it allows to visualize the spectrum contained in each spaxel.}, the science extension of the cube must be copied into a new file, e.g.: 

\begin{lstlisting}
imcopy dstexrgS20110923S0268.fits[sci,1]  star.fits
\end{lstlisting}

\clearpage{\pagestyle{empty}\cleardoublepage}

\chapter{The Science Data}			         
\label{chap: science}
With ``science data" we indicate images/spectra of the main target (galaxy, star cluster, \ldots) which will be studied to achieve the intended scientific goal. 

At this point we will suppose that the calibration star was successfully reduced and any problems with fiber misidentification and the mdf were fixed (see appendix \ref{app1}). If data were acquired on different days, it is recommended to separate them according to the acquisition date, reduce them separately and combine the final products at the end of the process.

\section{Prepare IRAF}			         
If running a new IRAF session, follow the steps outlined in \textsection\ref{sec: prepareIRAF} and define a new log file:

\begin{lstlisting}
gmos.logfile=``Target_Date.log''
\end{lstlisting}

If a user-customized mdf is to be used, then inform $gfreduce$ (let's assume the new mdf is located in the folder with the processed files):
\begin{lstlisting}
gfreduce.mdfdir = ``proc$''
gfreduce.mdffile = ``gsifu_slits_mdf_modified.fits'' 
\end{lstlisting}

\section{Flats reduction}	
Here we reduce the flats acquired with the science data. If the observations of both the star and the science data were performed in the same night then it is likely that these flats can be reduced in non-interactive mode. Otherwise, it is recommended to reduce them in interactive mode, define a new reference file and a new response curve.
 
\begin{lstlisting}
gfreduce @science_flat_list
fl_gscrrej- \
fl_wavtran- \
fl_skysub- \
fl_inter- \
fl_over- \
weights=none \
slits=both
\end{lstlisting}

Note 1: to apply the reduction step to multiple files simultaneously, inputs can be given in a list or by specifying each name separated by coma without spaces. Lists can be created with $gemlist$.

Note 2: if the bias was overscan subtracted, in this step the overscan should also be subtracted from the science data.

\section{Reduction of the arc}
In this case all of the observations were made on the same night, so the reference file will be the first flat which was already reduced in interactive mode for the standard star (ergS20110923S0252). As specified in the previous step, if observations were carried out over multiple nights, then it is recommended to interactively reduce the flats obtained with the science data and to define a new reference file.
 
 \begin{lstlisting}
gfreduce S20110923S0300 \
fl_wavtran- \
fl_inter- \
ref=ergS20110923S0252 \
recenter- \
trace- \
fl_skysub- \
fl_gscrrej- \
fl_bias- \
fl_over- \
order=1 \
weights=none \
slits=both
 \end{lstlisting}
 
Note 1: if spectral dithering was required, there will be more than one arc that needs to be reduced.
 
Note 2: if the bias was overscan subtracted, in this step the overscan should be subtracted from the arc.
 
Note 3: if a new folder for the processed data was defined for the science data and the reference file from standard star is used, then IRAF will give an error while extracting the slits: \textit{Extracting slit 1, ERROR - GFEXTRACT: Aperture reference database/aperg... not found}. This is because IRAF created a subfolder of ``proc" (the folder with the processed files) called ``database" with the reference files. 
 
 This is an example of why it is easier to work in two directories only: $raw$ and $proc$.

\section{Establish wavelength calibration}

See \textsection \ref{sec: establish_wc} for comments.
 \begin{lstlisting}
gswavelength ergS20110923S0300 \
fl_inter+ \
nlost=10 \
order=3
 \end{lstlisting}
 
Note 1: if spectral dithering was requested, there will be more than one arc, and this operation needs to be performed separately on each arc.
 
Note 2: if observations have been performed over multiple nights, there will be more than one arc, and this operation needs to be performed separately on each arc.

\section{Apply wavelength calibration on the arc}
 See \textsection\ref{sec: apply_wc_arc} for comments.
 
 \begin{lstlisting}
 gftransform ergS20110923S0300 \
 wavtran=ergS20110923S0300
  \end{lstlisting}

\section{Reduction of the science data}
\subsection{Bias subtraction} 

Here we subtract the bias from the science data and trim them. Using \textit{gemlist} we may create lists of images for each spectral dither and run gfreduce:

 \begin{lstlisting}
 gfreduce @Target_filter1_list \
slits=both \
fl_inter- \
fl_addmdf+ \
key_mdf=MDF \
fl_over- \
fl_trim+ \
fl_bias+ \
fl_flux- \
fl_gscrrej- \
fl_extract- \
fl_gsappwave- \
fl_wavtran- \
fl_skysub- \
weights=none
 \end{lstlisting}
 
 \ldots and so on for the other filters. In preparation for the next step, make a list of the resulting images:
 
 \begin{lstlisting}
 gemlist rgS20110923S 249-251,257-259 > rgTarget_filter1_list 
 gemlist ... 	> rgTarget_filter2_list 
 \end{lstlisting}

\subsection{Flatfielding and spectra extraction} 	
Extract the spectra, apply the flat field correction and remove cosmic rays from the science data. 
  
\begin{lstlisting}
gfreduce @rgTarget_filter1_list \
fl_inter- \
fl_addmdf- \
fl_over- \
fl_trim- \
fl_bias- \
fl_gscrrej+ \
fl_extract+ \
fl_wavtran- \
fl_sky- \
fl_flux- \
slit=both \
trace- \
verb+ \
refer=ergS20110923S0269 \
response=ergS20110923S0269_resp \
weights=none
 \end{lstlisting}

Note: when using \textit{fl\_gscrrej+} for cosmic ray removal it is advised to check if IRAF also removes features not due to cosmic rays. 
 
If convenient, make a list of the resulting files to apply the next step:
 
 \begin{lstlisting}
 gemlist exrgS20110923S 249-251, ... > exrgTarget_filter1_list 
 \end{lstlisting}

\subsection{Wavelength calibration}	 		
Using the appropriate processed arc for each spectral dither (in this case \textit{ergS20110923S0299}), apply the wavelength calibration:
 \begin{lstlisting}
gftransform @exrgTarget_650_list \
wavtran=ergS20110923S0299
 \end{lstlisting}

\ldots and make lists to proceed to the next step:

\begin{lstlisting}
gemlist texrgS20110923S 249-251, ... > texrgTarget_filter1_list 
gemlist  ... > texrgTarget_filter2_list 
\end{lstlisting}

\subsection{Sky subtraction}			 
 For each spectral dither: 
 
 \begin{lstlisting}
gfreduce @texrgTargetName_filter1_list  \
fl_inter- \
fl_addmdf- \
fl_over- \
fl_trim- \
fl_bias- \
fl_gscrrej- \
fl_extract- \
fl_wavtran- \
fl_sky+ \
fl_flux- \
slit=both \
trace- \
verb+ \
weights=none
 \end{lstlisting}
 
  Output prefix: ``s".
 
 Make lists with results:
 \begin{lstlisting}
gemlist stexrgS20110923S 249-251, ... > stexrgTarget_filter1_list 
gemlist ... > stexrgTarget_filter2_list 
 \end{lstlisting}

\subsection{Flux calibration}			 		
Now we use the sensitivity function created from the star. For each filter:
 
 \begin{lstlisting}
 gscalibrate @stexrgTargetName_filter1_list \
 sfunction=``sens.fits"\
 observa=``Gemini-South"\
 fluxscal=1
 \end{lstlisting}
 
 Output prefix: ``c".

\subsection{Creating the cube}		 		
\label{sec: cubes}
To create a 3D (x, y, $\lambda$) datacube apply \textit{gfcube} to each processed science image:
\begin{lstlisting}
gfcube cstexrgS20110923S0249 ssample=0.05
gfcube cstexrgS20110923S0250 ssample=0.05
...
\end{lstlisting} 
 
where \textit{ssample} indicates the spatial sampling, in this case 0.05 $arcsec/pxl$. Note that this value should not be greater than the FWHM of the seeing.  Output prefix: ``d".
In order to visualize the cube with, e.g., QFitsView the science extension of the cube must be copied into a new image:
 
 \begin{lstlisting}
imcopy dcstexrg...249.fits[sci,1] dcstexrg...249_sci.fits 
imcopy dcstexrg...250.fits[sci,1] dcstexrg...250_sci.fits
...
 \end{lstlisting}
 
At this point we can combine the cubes containing the science images.

\subsubsection{Using imcombine}
We have to specify spatial and spectral offsets between one image and the other (offsets are usually required by the principal investigator of the observation; spatial offsets are necessary to remove cosmic rays and the effects of dead fibers; spectral offsets are used also for cosmic rays and to fill the three gaps in the spectrum due to the GMOS detector). 
  
Spatial offsets (in arcseconds) are specified in the headers of the files $dcstexrg$\ldots\ under the keywords \textit{xoffset} and \textit{yoffset}.
Offsets must be converted to pixels (using the pixel scale) and written to a text file with 3 columns where the first is the \textit{xoffset}, the second is the \textit{yoffset} and the third is the offset in the spectral dimension (in angstroms). As a sanity check, it is useful to visually inspect the images to be combined. In particular, visual inspection will help to determine that signs (+ or -) are correct (e.g. if the galaxy is on top of the frame this means that the telescope was offset downward: this corresponds to a negative offset along the $y$ direction). Once the file with the offsets is created, we can make a list with all of the science files to combine and we can proceed to the final step:

\begin{lstlisting}
imcombine @files.list OutputName \
combine= ``average''\
reject=``avsigclip'' \
project- \
outtype=``real'' \
offsets=``offsets''
 \end{lstlisting}

This process takes up to 45 minutes to combine 12 cubes on a 2.53 GHz Intel Core 2 Duo machine. 

Once the cube is created we can check the result using QFitsView. If the image looks like Fig.\ref{fig: wrong_offsets}, this means the offsets were not taken into account correctly. In this case, it is necessary to go back, check the combined files and the offsets specified in the file offsets.txt and given in input to \textit{imcombine}.

\subsubsection{Using gemcube}
Gemcube does not require an input file with the offsets, nevertheless keywords like \textit{crval}, \textit{crpix}, \textit{cd} need to be properly added to the header file. 

In this example we use the continuum peak (assumed to be visible in all the cubes) as a reference point to align the cubes and define a new coordinate system centered on it. The pixel corresponding to the new center is specified by the header keywords CRPIX1 and CRPIX2. Let's assume that XCP and YCP are the coordinates of the continuum peak measured on the cube \textit{dcstexrgS20110923S0249\_sci.fits}. We will choose this point as the origin of the new coordinate system by setting (CRVAL1, CRVAL2) = 0.

\begin{lstlisting}
hedit dcstexrg...249_sci.fits CRPIX1 XCP add+ verify- update+
hedit dcstexrg...249_sci.fits CRVAL1 0. add+ verify- update+
hedit dcstexrg...249_sci.fits CRPIX2 YCP add+ verify- update+
hedit dcstexrg...249_sci.fits CRVAL2 0. add+ verify- update+
\end{lstlisting}

Now we store the information regarding the pixel size (set with $gfcube$ at the begin of this section): 
\begin{lstlisting}
hedit dcstexrg...249_sci.fits CD1_1  0.0 add+ verify- update+
hedit dcstexrg...249_sci.fits CD1_2  -0.05 add+ verify- update+
hedit dcstexrg...249_sci.fits CD2_1  -0.05 add+ verify- update+
hedit dcstexrg...249_sci.fits CD2_2  0.0 add+ verify- update+
\end{lstlisting}
Having CD1\_1 and CD2\_2 equal zero also implies that the image is oriented with North up. Though this is not necessarily true (very likely it isn't). The actual orientation of the image can be found in the raw data. 

Finally let's set:
\begin{lstlisting}
hedit dcstexrg...249_sci.fits WAT1_001 'linear' add+ verify- update+
hedit dcstexrg...249_sci.fits WAT2_001 'linear' add+ verify- update+
hedit dcstexrg...249_sci.fits CTYPE1 'linear' add+ verify- update+
hedit dcstexrg...249_sci.fits CTYPE2 'linear' add+ verify- update+
\end{lstlisting}

These steps must be repeated for all cubes that will be combined (obviously XCP and YCP will be different for each cube). Once all of this is done we can make a list of the files to combine and merge them to create the cube:

\begin{lstlisting}
gemcube @list.txt Target.fits
\end{lstlisting}

\subsubsection{Which one is better? Imcombine or gemcube?}
I have been suggested to use of \textit{imcombine} because it works with an efficient rejection algorithm for bad pixel/cosmic rays removal (\textit{sigclip}, \textit{avsigclip}), nevertheless sometimes I obtain better results with one and sometimes with the other.

\clearpage{\pagestyle{empty}\cleardoublepage}

		\renewcommand{\sectionmark}[1]{\markright{\thesection\ #1}}  
	
\clearpage{\pagestyle{empty}\cleardoublepage}

\appendix

\chapter{Typical problems}			
\label{app1}
\label{subsubsec: problems}

\section{Fibers Misidentification}
\begin{figure}[p]
\centering
\includegraphics[trim = 0cm 0cm 0.15cm 0cm, clip = true, draft=false, scale=0.29]{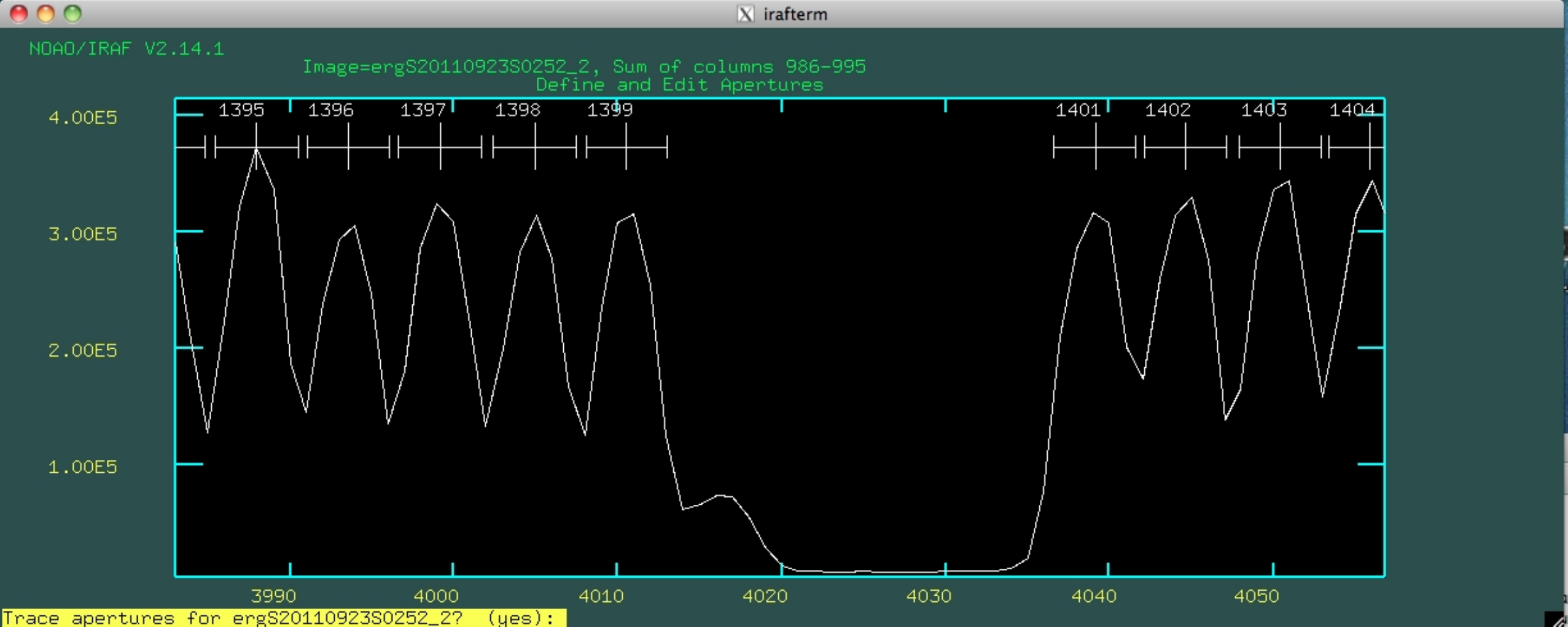} 
\caption[Example of a missing fiber]{A missing fiber between a fiber group and the next group. Sometimes the missed fiber is totally invisible. Note the IDs on top.} 
\label{fig: missing_fiber}
\end{figure}

It seems that with GEMINI-South fibers misidentification is more common than with GEMIN-North. We may find these two problems:
\begin{enumerate}
\item Missing fibers. Suppose to represent a set of fibers as 1 1 0 1, where 1 is the good fiber and 0 is the bad fiber, so we should have the ID assigned as, e.g.: 4 5 empty 7. If we have: 4 5 empty 6, this would be wrong and the mdf must be modified. First we copy the mdf from the IRAF/GEMINI database into the directory with the processed files: 

\begin{lstlisting}
  copy gmos$data/gnifu_slits_mdf.fits .
\end{lstlisting}

(include the ``.'' at the end of the command). Now we modify the copied mdf:

\begin{lstlisting} 
  tcalc gnifu_slits_mdf.fits BEAM ``if NO == xxx 
  then -1 else BEAM''
\end{lstlisting}

where xxx is the number of the missing fiber. In our example this is 6. 

Finally, we must inform IRAF about the location of the new mdf:

\begin{lstlisting} 
  gfreduce.mdfdir = ``proc$''
\end{lstlisting}

See also: \url{http://ifs.wikidot.com/gmos}.

\item IRAF identifies the same fiber twice. We have to correct the misidentification manually: as IRAF assigned two IDs to the same fiber, there are two numbers superposed in the same location of the interactive window. After placing the red cross on the fiber, the command ``d" will delete the ID. IRAF will ask which fiber ID should be deleted, e.g. 137 or 139? Delete the 139th and type ``r" to refresh the plot. At this point we could ask IRAF to identify the fibers again by hitting ``f", possibly after deleting and marking with ``m" other fibers\footnote{If using ``m", IRAF will center the fiber position automatically, using ``n" the fiber position will be set in the point where we put the marking cross}. Sometimes this process does not work. In that case, we have to delete all remaining fibers ID and re-mark them by hand (it requires some experience and it is tedious, yet interactive identification of the fibers could be the best approach to be used - always). 

It may happen that between one block and the other the gap is unusually large, see Fig.\ref{fig: missing_fiber}. This is due to a missing fiber just before/after the gap. So, mark the last line of the block, let's say it is xx9, then move to the next block and mark the first line. If the number is xx1, the identification is working properly (this also means that such a fiber was already marked as a missing fiber into the mdf). Note: sometimes, at the beginning, or between one gap and the other, there is a bump. IRAF might identify it as a fiber. It must be deleted and the following fiber must be marked.

NOTE: if this problem is not fixed, all references will be wrong and the final output will be useless. If fiber misidentification becomes apparent at some point during the reduction process, we can go back and run:

\begin{lstlisting} 
 gfextract rg...filename fl_inter+
 \end{lstlisting} 
 or we can run \textit{gfreduce} again on the raw flat (this takes more time because it calls additional routines).
Once fibers are correctly identified, all of the processes calling a reference file must be run again.
 
\end{enumerate}

\begin{figure}[p]
\centering$
\begin{array}{cc}
\includegraphics[trim = 13cm 0cm 13cm 1cm, clip = true, draft=false,  scale=0.3]{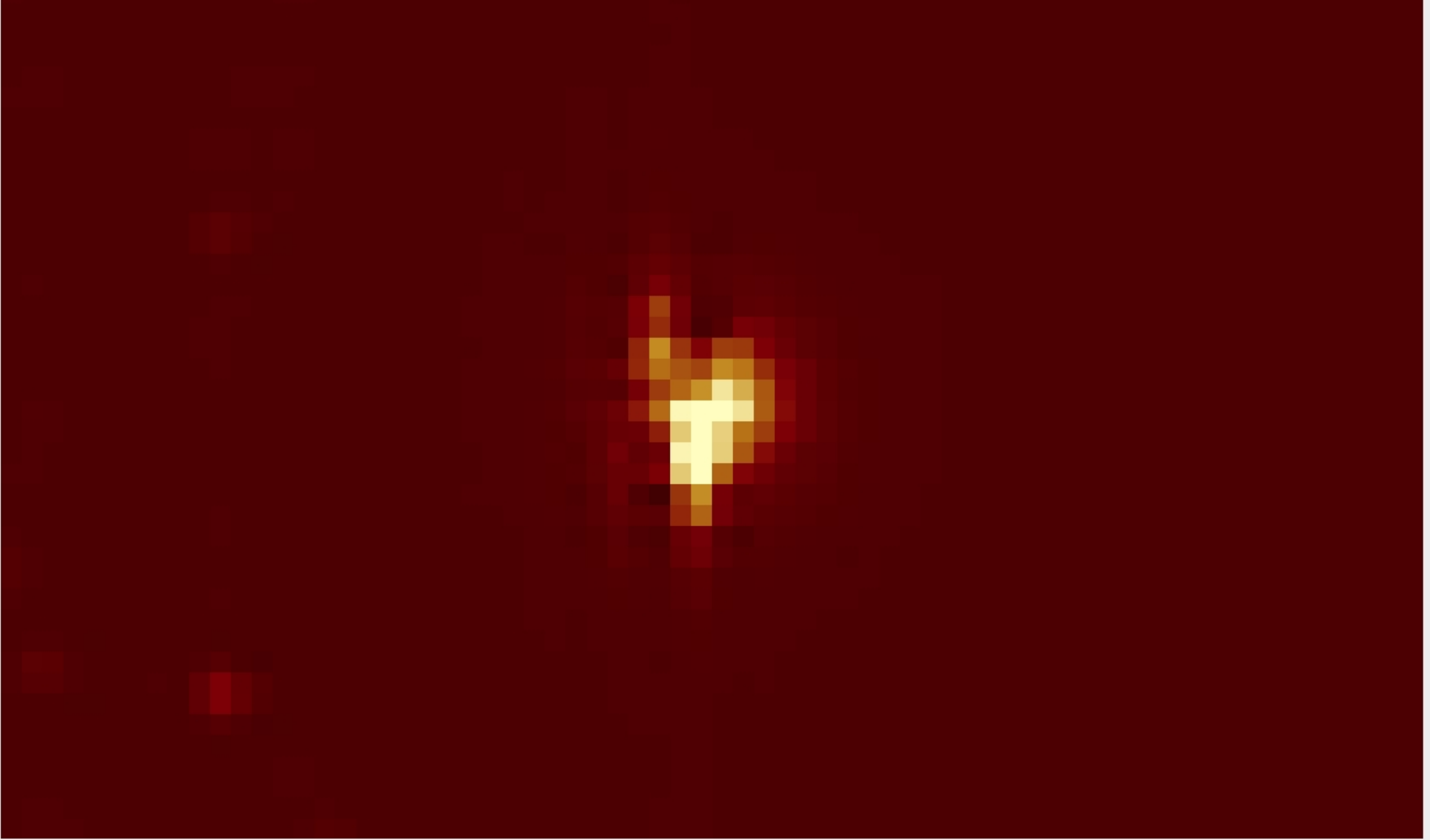} & \includegraphics[draft=false,  scale=0.3]{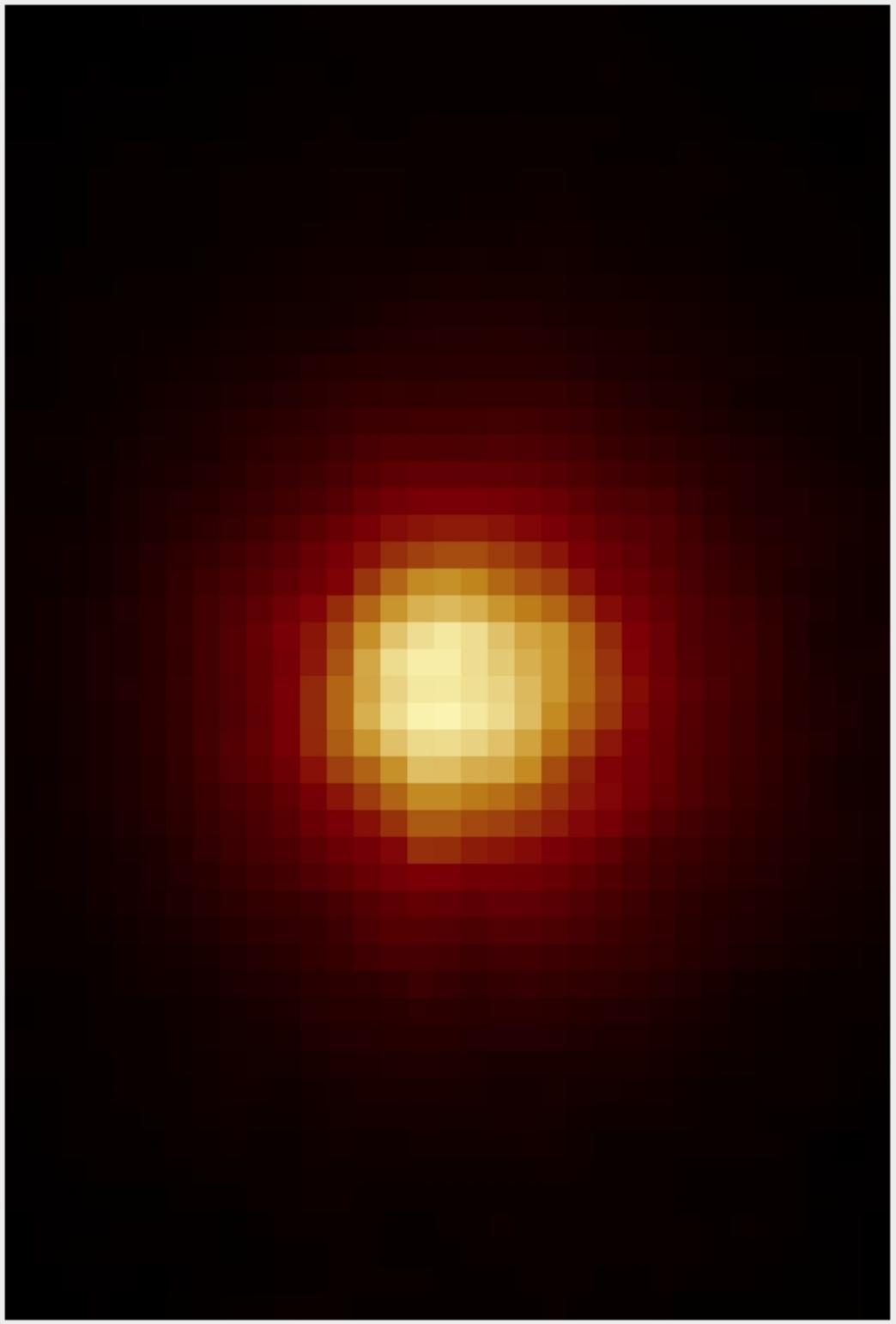} 
\end{array}$
\caption[Result of fibers misidentification]{Reconstructed image of a star where fibers were misidentified (left) and correctly identified (right).} 
\label{fig: mis_fibers}
\end{figure}

\section{Incorrect sensitivity function}
\label{app: SF}
\begin{figure}[p]
\centering
\includegraphics[trim = 0cm 0.125cm 0.125cm 0cm, clip=true, draft=false, width= \textwidth]{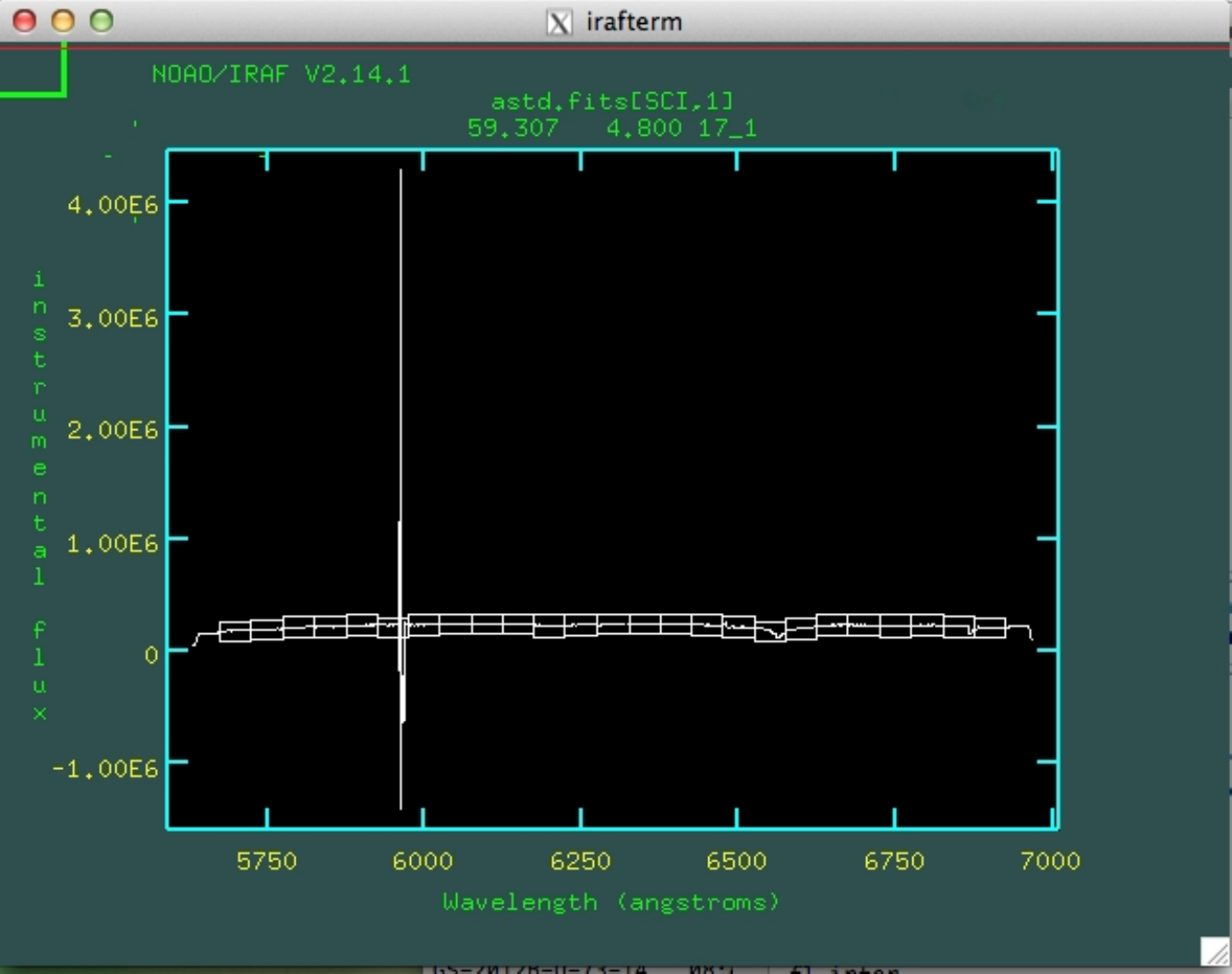}
\caption[Bad feature in instrumental flux vs wavelength]{Instrumental flux as a function of wavelength. The spikes at $\lambda < 6000$ \AA \ are bad features that should have been flagged by the rejection algorithm used in \textit{gfapsum}. If all rejection algorithms fail, the flagging may be performed interactively using the key ``d" on the bandpasses to flag. The corrected version of this spectrum is showed in Fig.\ref{fig: bandpasses_correct}.} 
\label{fig: bandpasses_problem}
\end{figure}

\begin{figure}[p]
\centering
\includegraphics[trim = 0.1cm 0cm 0.1cm 0cm, clip = true, draft=false, width= \textwidth]{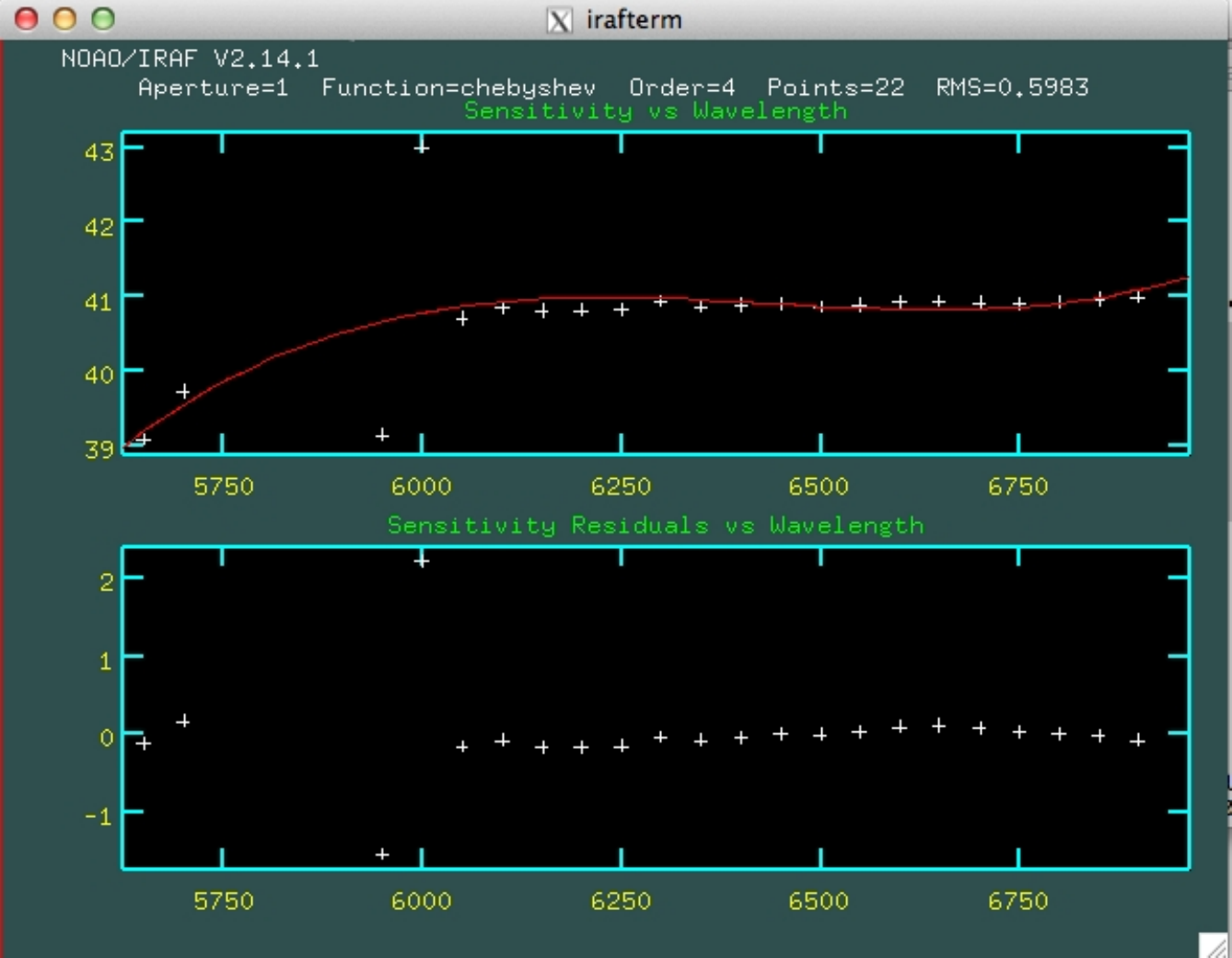}
\caption[Bad feature in the sensitivity curve]{Sensitivity curve. The effect of the unflagged spikes shown in Fig.\ref{fig: bandpasses_problem} is clearly visible at $\lambda < 6000$ \AA\ where a number of points are out of scale and the few visible points are likely to be affected as well. The visible points can be excluded from the fit using the stroke ``d". A new point may be added using the key ``a" and the fit may be run again using ``f". Note: before using this interactive (and somewhat subjective method) explore other rejection algorithms in \textit{gfapsum}.} 
\label{fig: sensitivity_wrong}
\end{figure}

In \textsection \ref{chap: star} we built the sensitivity function (SF). This was a pivotal step in the reduction process because the SF is used to calibrate the flux of the main astronomical object. If there is a problem in the SF, there will be a problem in the flux of the main object. Let see an example: in \textsection \ref{subs: sensitivity} we summed all spectra for the standard star using the following commands:

 \begin{lstlisting}
gfapsum stexrgS20110923S0268 \
outimages=``astd'' \
combine=``sum''\
reject=``pclip''\
fl_inter-
 \end{lstlisting}

and everything was fine (at least, for me). Yet, in my particular case, a different rejection algorithm, namely \textit{reject=``avsigclip''}, failed to detect some features that should have been flagged. Let's see the result. Suppose we used \textit{avsigclip}. We move forward in the process of creating the sensitivity curve with:

\begin{lstlisting}
gsstandard astd \
sfile=``std3'' \
sfunctio=``sens3'' \
starnam=``L1788'' \
fl_inte+ \
observa=``Gemini-South'' \
functio=``chebyshev'' \
order=4 \
\end{lstlisting}

IRAF will ask if we want to edit the bandpasses. We accept and Fig.\ref{fig: bandpasses_problem} appears. This is much different than the spectrum displayed in the case with no problems (Fig.\ref{fig: bandpasses_correct}). At $\lambda < 6000$ \AA\ there are spikes that were not flagged by the aforementioned rejection algorithm (\textit{avsigclip}). If the spikes are not flagged, either by the rejection algorithm or interactively (at this step or at the following one), a sensitivity curve similar to the one showed in Fig.\ref{fig: sensitivity_wrong} will result. Note that a number of points are completely out of scale for $\lambda < 6000$ \AA.
If we opted for an interactive fit, the bad points may be flagged and substituted with the keyboard strokes ``d" and ``a". A new fit has to be run with ``f".

The interactive flagging may somewhat fix the sensitivity curve, leaving room for more or less educated/reasonable subjectivity in the replacement of the points. Therefore, instead of choosing this solution, it is strongly recommended to go back, change the rejection algorithm and see if this fixes the problem. 

\begin{figure*}[p]
\begin{center}$
\begin{array}{cc}
\includegraphics[scale=0.515]{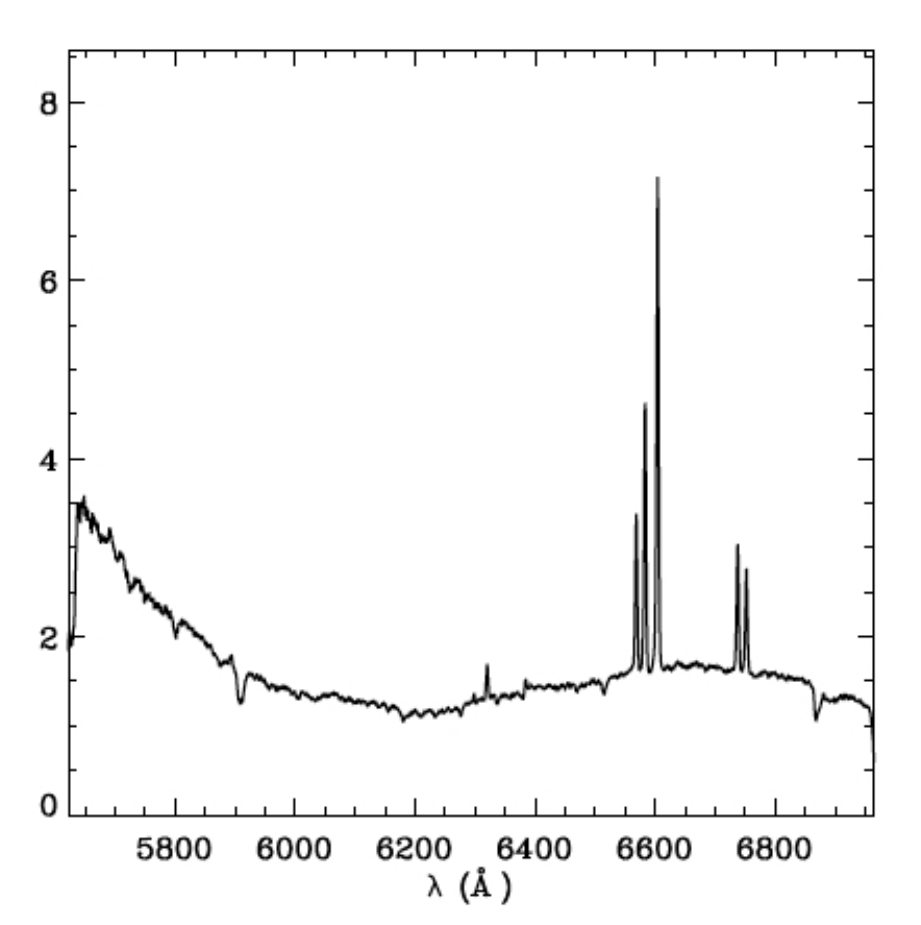} \\
 \includegraphics[trim=0cm 0cm 0cm 0cm, clip=true, scale=0.33]{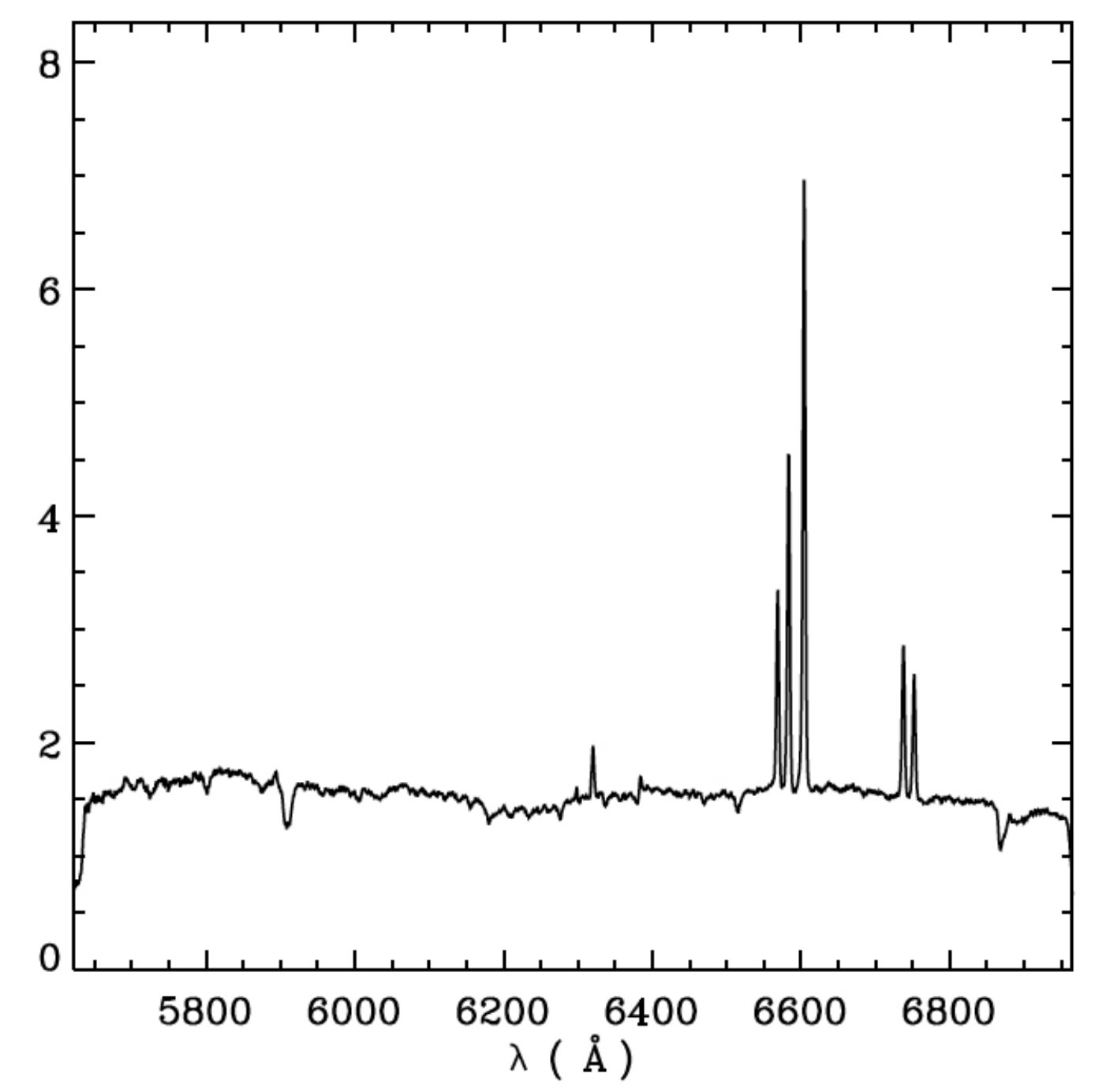}\\
\end{array}$
\end{center}
\caption[Spectra obtained with bad and good sensitivity function]{Spectra obtained using the wrong (top) and the correct (bottom) sensitivity function. Note the strong flux increase at short wavelengths caused by the decrease of the sensitivity function clearly visible in Fig.\ref{fig: sensitivity_wrong}. }
 \label{fig: subr_spec}
\end{figure*}

Fig.\ref{fig: subr_spec} shows the effect that such a sensitivity curve would have on the final spectrum of the galaxy.
\vskip10pt

\textbf{Final remark:} the presence of the large-aplitude spike near $\lambda = 6000$ \AA\ seems to be present in many (all?) observations obtained in the period 2010-2011 with GMOS-South.

\section{Offsets not properly considered}
\begin{figure}[p]
\centering$
\begin{array}{cc}
\includegraphics[draft=false,  scale=0.55]{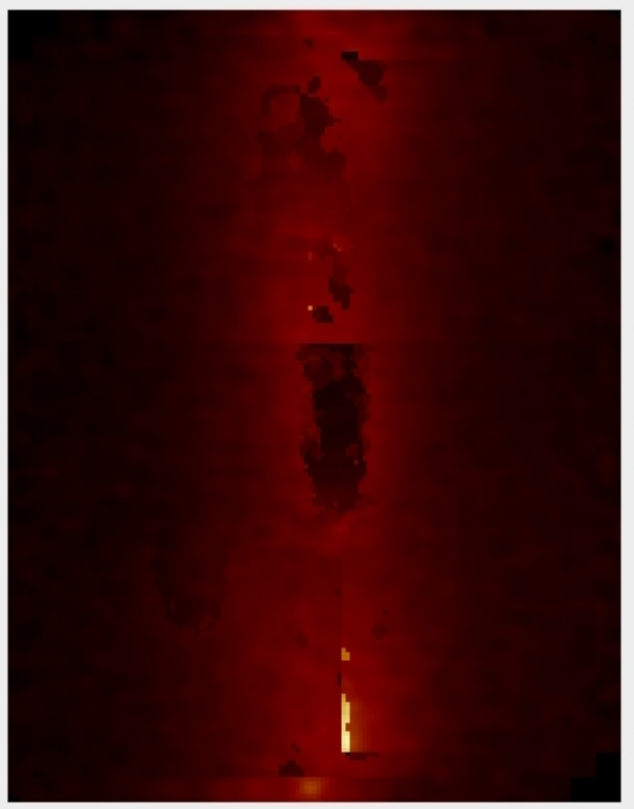} & \includegraphics[scale=0.55]{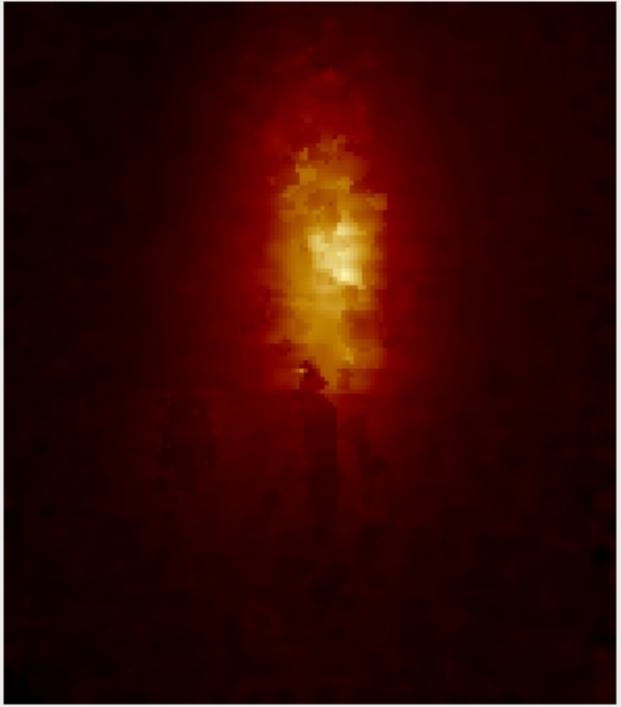} 
\end{array}$
\caption[Reconstructed images using wrong offsets]{Results of combinations of science images where the offsets between the different exposures have not been taken properly into account.} 
\label{fig: wrong_offsets}
\end{figure}

Usually the PI requires to take images of the main target offsetting the telescope. This is mainly to take care of dead fibers: if all images are taken with the same configuration, a dead fiber will result in a ``hole" in the final image. When combining the images in the very last steps of the reduction process, \textsection \ref{sec: cubes}, these offsets have to be properly given in input to \textit{imcombine}, otherwise the resulting image may look like Fig.\ref{fig: wrong_offsets}.

\clearpage{\pagestyle{empty}\cleardoublepage}

\section{Poor wavelength calibration}
\label{sec: poor_wc}
\begin{figure}[h]
\centering$
\begin{array}{cc}
\includegraphics[trim=16cm 2cm 11cm 0cm, clip=true, scale=0.32]{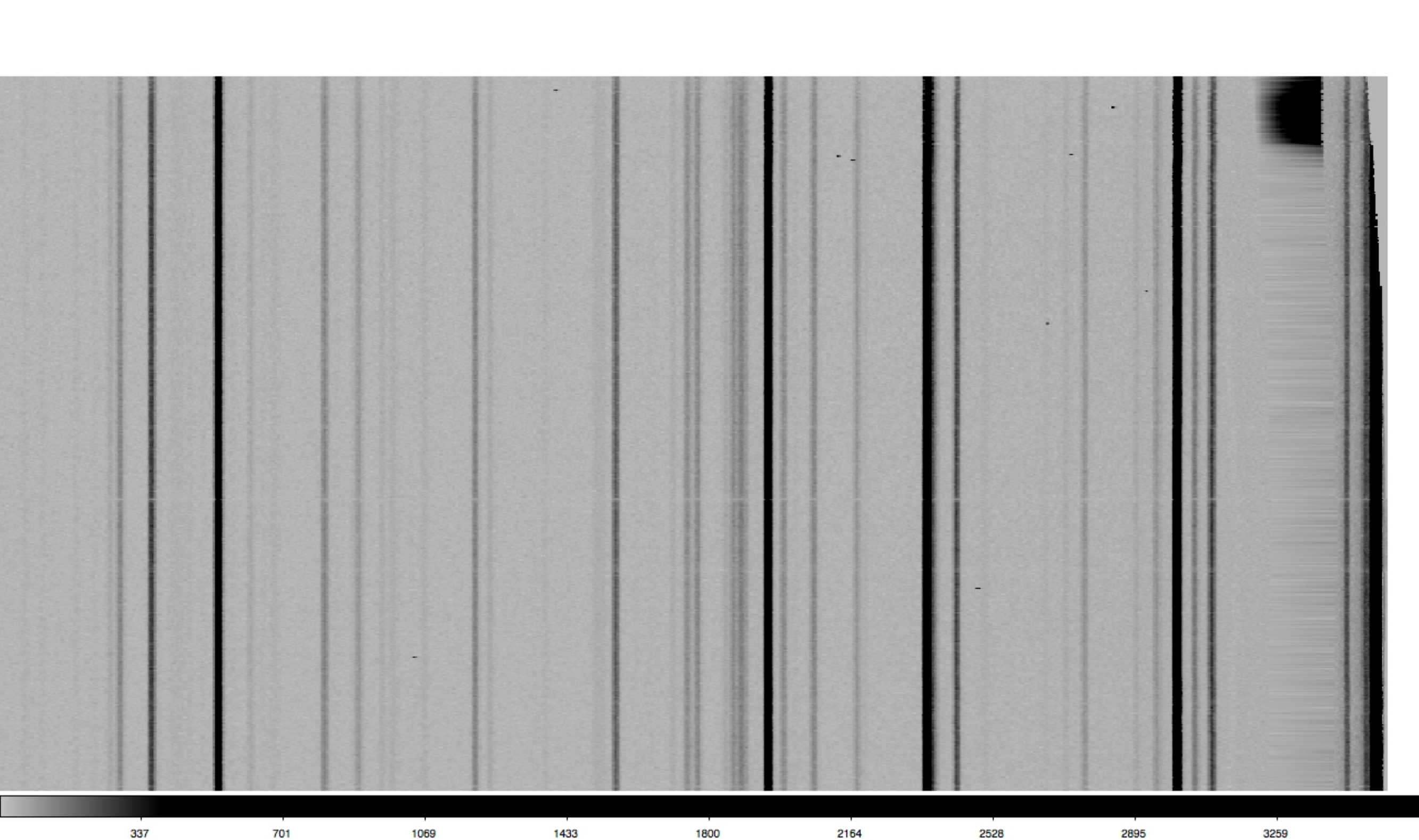} & \includegraphics[trim=16cm 2cm 11cm 0cm, clip=true, scale=0.32]{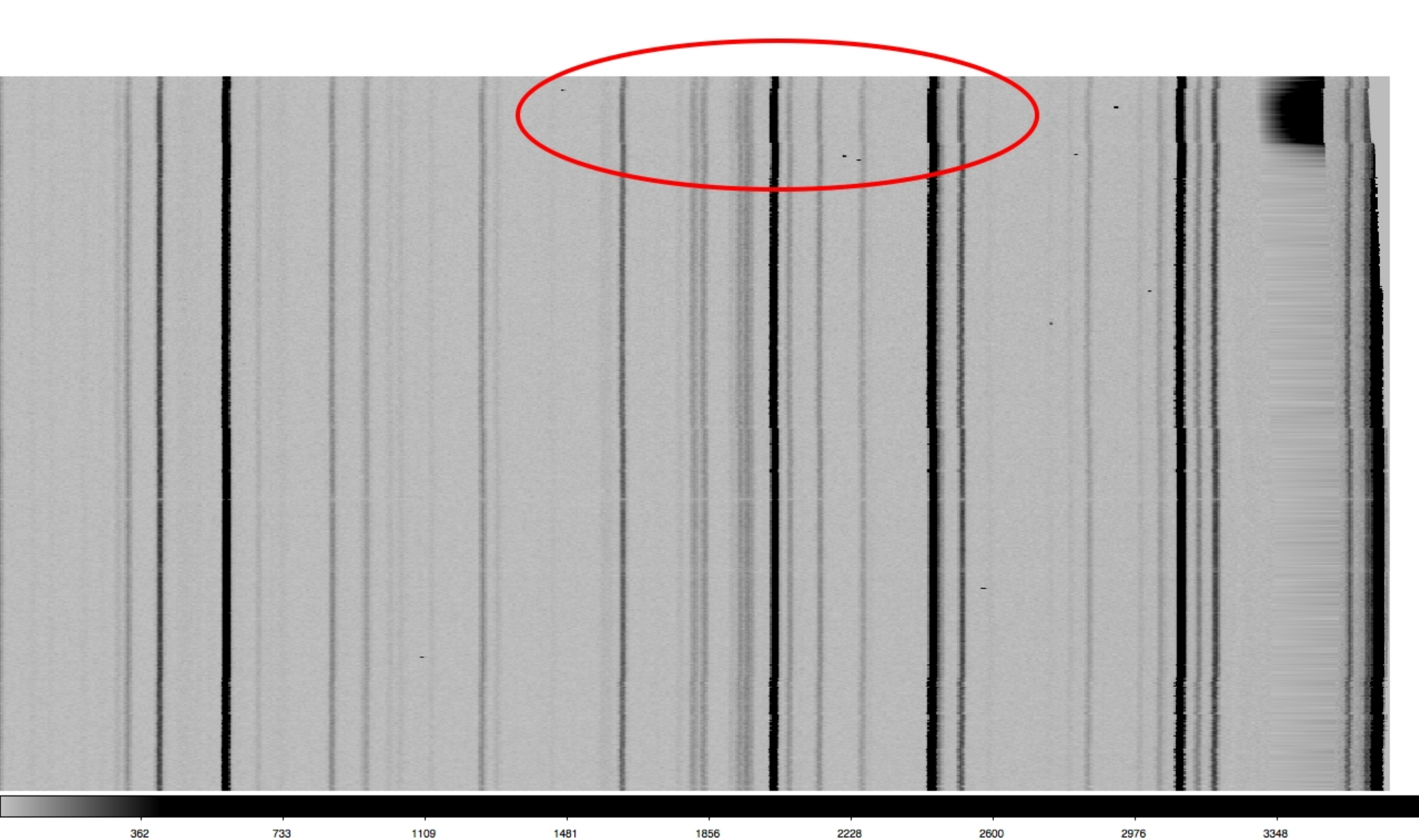} 
\end{array}$
\caption[Good and poor wavelength calibration]{Examples of wavelength calibrated arcs. \textit{Left}: good calibration obtained by fitting a third order chebyshev polynomial. \textit{Right}: poor calibration obtained by fitting a fourth order chebyshev polynomial. In the right image lines look broken in several points (as in the highlighted region).} 
\label{fig: good_and_poor_wc}
\end{figure}

If the function used to determine the wavelength calibration has an excessively high order (with respect to what is allowed by the quality of the data to fit), then it could fit features that should not be fitted, producing a poor wavelength calibration. Fig.\ref{fig: good_and_poor_wc} shows a wavelength calibrated arc using a third and fourth order chebyshev polynomial in \textit{gswavelength}. The result of the features shown in Fig.\ref{fig: good_and_poor_wc} is that spectra in different pixels of the science data-cube will artificially shift in wavelength (it would be as if different wavelength calibrations would have been applied to different pixels).

\chapter{Useful Definitions}
\label{app: def}
\begin{figure}[p]
\centering
\includegraphics[draft=false,  width= \textwidth]{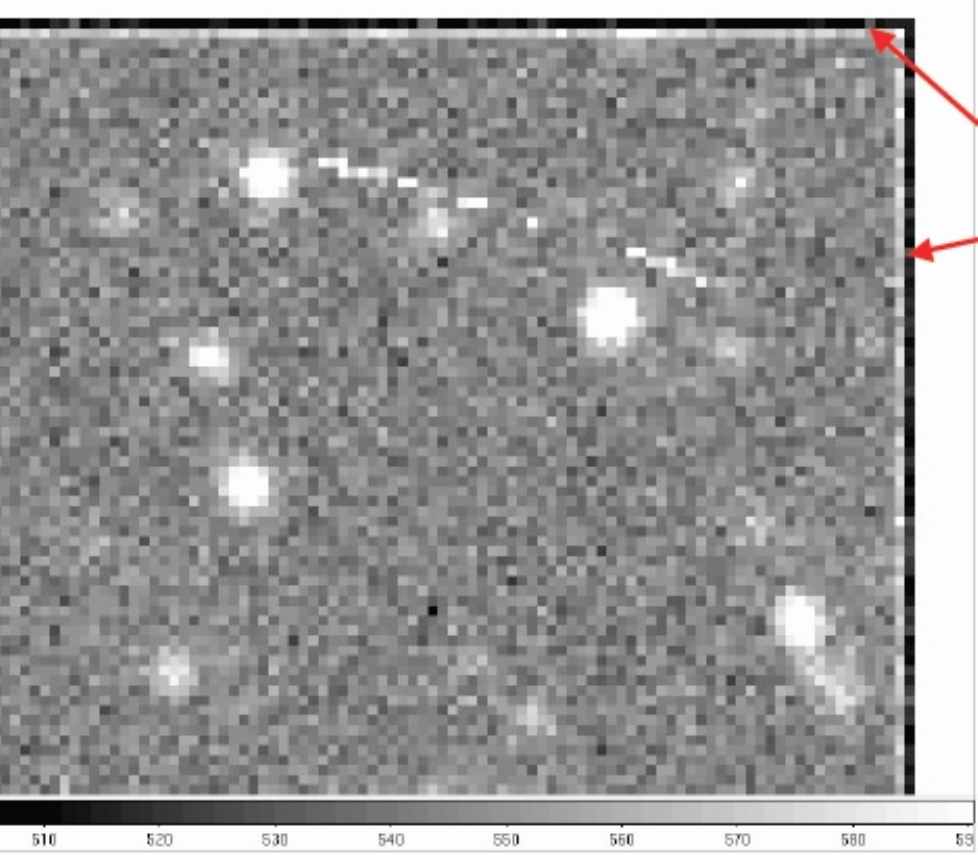}
\caption[Overscan regions]{Overscan regions indicated by red arrows. Credits: \url{http://www.astro.virginia.edu/~odf4n/snirs/reduction_1.pdf}} 
\label{fig: overscan}
\end{figure}

\begin{description}
\item[Lenslet arrays] are arrays of small lenses. GMOS has hexagonal lenslets that fully sample the field of view (FOV). E.g. the ``two slits mode" object FOV, $5 \times 7$ arcsec, has 1000 lenslets.
\item[MDF] or Mask Definition File is a FITS file which specifies which fibers are good and which fibers are missing. A mask definition file can be edited as a table using the IRAF command \textit{tedit}. Once opened with \textit{tedit}, note the column BEAM: values equal to 1 indicate the presence of a good aperture, -1 indicate the presence of a missing fiber.
\item[Overscan regions] are columns and rows not exposed to light along the edges of an image, Fig.\ref{fig: overscan}. Note: the overscan is taken at the same time of the data, hence it reflects actual electronic noise and time variability. However, the overscan does not contain the full 2D informations contained in a dark or bias.
\item[``Tracing fibers"] means to assign an ID to each fiber. 
\end{description}

\chapter{Suggested readings}
\label{sec: reading}
\begin{itemize}
\item \href{http://adsabs.harvard.edu/abs/2002PASP..114..892A}{``Integral Field Spectroscopy with the Gemini Multiobject Spectrograph. I. Design, Construction, and Testing"} Allington-Smith, Jeremy; Graham, M.; Content, R.; Dodsworth, G.; Davies, R.; Miller, B. W.; Jorgensen, I.; Hook, I.; Crampton, D.; Murowinski, R., (2002) PASP, 114, p. 892-912

\item \href{http://adsabs.harvard.edu/abs/2004PASP..116..425H}{``The Gemini-North Multiobject Spectrograph: Performance in Imaging, Long-slit, and Multi-Object Spectroscopic Modes"}, Hook, Isobel; J¿rgensen, Inger; Allington-Smith, J. R.; Davies, R. L.; Metcalfe, N.; Murowinski, R. G.; Crampton, D., (2004), PASP, 116, p. 425-440
\end{itemize}

	\fancyhead[RE]{\bfseries\footnotesize\nouppercase{\leftmark}}
	\fancyhead[LO]{\bfseries\footnotesize\nouppercase{\rightmark}}

\clearpage{\pagestyle{empty}\cleardoublepage}

\end{document}